\documentclass[aps,prd,onecolumn,groupedaddress,showpacs,nofootinbib,amssymb]{revtex4}
\usepackage[dvips]{graphicx}
\usepackage{amssymb}
\usepackage{amsmath}
\usepackage{graphicx}
\usepackage{amsfonts}
\usepackage{bm,color}
\usepackage{amsmath,amssymb,theorem}
\newcommand\be{\begin{equation}}
\newcommand\ee{\end{equation}}
\newcommand\nn{\nonumber \\}
\newcommand\e{\mathrm{e}}

\allowdisplaybreaks[4]

\begin{document}

\tolerance=5000

\title{Ghost-free Gauss-Bonnet Theories of Gravity}
\author{S.~Nojiri,$^{1,2}$\,\thanks{nojiri@gravity.phys.nagoya-u.ac.jp}
S.~D.~Odintsov,$^{3,4}$\,\thanks{odintsov@ieec.uab.es}
V.~K.~Oikonomou,$^{5,6,7}$\,\thanks{v.k.oikonomou1979@gmail.com}}

\affiliation{$^{1)}$ Department of Physics, Nagoya University,
Nagoya 464-8602, Japan \\
$^{2)}$ Kobayashi-Maskawa Institute for the Origin of Particles
and
the Universe, Nagoya University, Nagoya 464-8602, Japan \\
$^{3)}$ ICREA, Passeig Luis Companys, 23, 08010 Barcelona, Spain\\
$^{4)}$ Institute of Space Sciences (IEEC-CSIC) C. Can Magrans
s/n,
08193 Barcelona, Spain\\
$^{5)}$ Department of Physics, Aristotle University of
Thessaloniki, Thessaloniki 54124,
Greece\\
$^{6)}$ Laboratory for Theoretical Cosmology, Tomsk State
University of Control Systems
and Radioelectronics, 634050 Tomsk, Russia (TUSUR)\\
$^{7)}$ Tomsk State Pedagogical University, 634061 Tomsk, Russia\\
}

\begin{abstract}
In this work we develop a theoretical framework for Gauss-Bonnet
modified gravity theories, in which ghost modes can be eliminated
at the equations of motion level. Particularly, after we present
how the ghosts can occur at the level of equations of motion, we
employ the Lagrange multipliers technique, and by means of
constraints we are able to eliminate the ghost modes from
Gauss-Bonnet theories of the form $f(\mathcal{G})$ and
$F(R,\mathcal{G})$ types. Some cosmological realizations in the
context of the ghost free $f(\mathcal{G})$ gravity are presented,
by using the reconstruction technique we developed. Finally, we
explore the modifications to the Newton law of gravity generated by
the ghost-free $f(\mathcal{G})$ theory.
\end{abstract}

\maketitle

\section{Introduction}

Undoubtedly one of the mysteries in theoretical physics is to find
a consistent way to describe all the observed interactions under
the same theoretical framework. This would require gravity to be
quantized in some way and up to date, only string theory seems to
provide a complete UV completion of all known particle physics
theories. In cosmology, the quantum gravity era controls the
pre-inflationary era, during which gravity is expected to be
unified with all the other three interactions. It is evident that
during this pre-inflationary era, string theory would be the most
appropriate theory to describe the physical laws of our Universe,
however it is not easy to prove that this is indeed the case.
However some string theory effects could have their impact on the
inflationary era, and this impact may be in fact measurable. There
exist many theories in modern theoretical cosmology which take
into account string theory motivated terms in the interaction
Lagrangian of the model, such as the scalar-Einstein-Gauss-Bonnet
gravity theory~\cite{Nojiri:2005vv,Nojiri:2006je}, in which case
the Lagrangian is of the form,
\begin{equation}
\label{FRGBg16}
S=\int d^4x\sqrt{-g} \left(\frac{1}{2\kappa^2}R
 - \frac{1}{2} \partial_\mu \chi \partial^\mu \chi
+ h\left( \chi \right) \mathcal{G} - V\left( \chi \right) +
\mathcal{L}_\mathrm{matter}\right)\, ,
\end{equation}
where $\mathcal{G}$ is the Gauss-Bonnet invariant defined as
follows,
\begin{equation}
\label{GB} \mathcal{G} \equiv R^2 - 4 R_{\mu\nu} R^{\mu\nu} +
R_{\mu\nu\rho\sigma} R^{\mu\nu\rho\sigma} \, .
\end{equation}
The scalar-Einstein-Gauss-Bonnet models are motivated by $\alpha'$
corrections in superstring theories~\cite{Gross:1986mw}, and they
serve as a consistent example of how string theory may leave its
impact on the primordial acceleration era of the Universe. Another
very well studied class of theories in the same context, is that
of $f(\mathcal{G})$
gravity~\cite{Nojiri:2005jg,Cognola:2006eg,Leith:2007bu,Li:2007jm,Kofinas:2014owa,Zhou:2009cy},
in which case the Lagrangian is of the form,
\begin{equation}
\label{GB1b} S=\int d^4x\sqrt{-g} \left(\frac{1}{2\kappa^2}R +
f(\mathcal{G}) + \mathcal{L}_\mathrm{matter}\right)\, .
\end{equation}
These theories contain a function of the Gauss-Bonnet invariant,
and therefore the presence of this function generates non-trivial
effects in the theory, due to the fact that the effect of the
Gauss-Bonnet term does not appear as a total derivative anymore,
as in the linear theory of the Gauss-Bonnet scalar. Both these
theories belong to a wider class of cosmological models which are
known as modified gravity models~\cite{Capozziello:2011et,Capozziello:2010zz,
Olmo:2011uz,
Nojiri:2017ncd,Nojiri:2010wj,Nojiri:2006ri,delaCruzDombriz:2012xy}, and
which generalize the standard Einstein-Hilbert theory. The
motivation for studying such theories comes from the fact that in
the context of these, several cosmological eras may be described
by the same theory in a unified way, see for example
Ref.~\cite{Nojiri:2003ft} in which the unified description of the
inflationary and of the dark energy eras was given in terms of
$f(R)$ gravity. In addition, similar studies were presented in
terms of scalar Einstein Gauss-Bonnet models~\cite{Odintsov:2018zhw}
and $f(\mathcal{G})$ models.

Due to the importance of the models containing or involving the
Gauss-Bonnet scalar, which are string theory motivated in most
cases, in this paper we shall address an important shortcoming of
these theories, namely the existence of ghosts. Usually,
higher-derivative theories contain ghost degrees of freedom due to
the Ostrogradsky's instability, see for example
\cite{Woodard:2015zca}. As was pointed out
in~\cite{DeFelice:2009ak}, ghost degrees of freedom may occur at
various levels of the theory, even at the cosmological
perturbations level of $F(R,\mathcal{G})$ theories, where
superluminal modes $\sim k^4$ occur, where $k$ is the associated
wavenumber. Having these issues in mind, in this paper we shall
investigate how the ghosts may be eliminated from $f(\mathcal{G})$
and $F(R,\mathcal{G})$ theories. Particularly, by using an
appropriate constraint used firstly in the context of mimetic
gravity~\cite{Chamseddine:2013kea,Nojiri:2014zqa,Dutta:2017fjw},
we shall demonstrate that the resulting theories are ghost-free.
Similar constrained Gauss-Bonnet theories in the context of
mimetic gravity were studied in~\cite{Astashenok:2015haa}. Also
ghost-free theories were also developed in
Refs.~\cite{Nojiri:2017ygt,Park:2018nfp}, but in a different
context. In this work we shall also consider the cosmological
evolution of the resulting theories, and we shall investigate how
several cosmological evolutions may be realized by the ghost-free
models we will develop, emphasizing on the dark energy era and
inflationary era. Finally, we shall investigate how the Newton law
is modified in the context of the ghost-free $f(\mathcal{G})$
gravity.

This paper is organized as follows: In section II we address the
ghost issue in the context of $f(\mathcal{G})$ gravity. We firstly
demonstrate how ghosts may occur in this theory and we provide two
remedy theories, which are ghost-free extensions of
$f(\mathcal{G})$ gravity. In section III we investigate how
several cosmological evolutions may be realized in the context of
the proposed ghost-free $f(\mathcal{G})$ theory. In section IV we
discuss how the Newton law becomes in the context of the
ghost-free $f(\mathcal{G})$ gravity, and finally in section V we
briefly investigate how a general $F(R,\mathcal{G})$ theory may be
rendered ghost free.

\section{Ghost-free $f\left( \mathcal{G} \right)$ gravity}

In this section we shall investigate how to obtain a ghost free
$f(\mathcal{G})$ gravity, and we shall employ the Lagrange
multipliers formalism in order to achieve this. Before getting
into the details of our formalism, we will start the presentation
by showing explicitly how ghost modes may occur in $f\left(
\mathcal{G} \right)$ gravity at the equations of motion level, and
the ghost-free version construction of the theory follows.

\subsection{Ghosts in $f\left( \mathcal{G} \right)$ Gravity}

In order to investigate if any ghost modes could appear in
$f\left( \mathcal{G} \right)$ gravity model (\ref{GB1b}), we
investigate the equations of motion, by considering a general
variation of the metric of the following form,
\begin{equation}
\label{Q3} g_{\mu\nu}\to g_{\mu\nu} + \delta g_{\mu\nu}\, .
\end{equation}
Effectively, the variations of $\delta\Gamma^\kappa_{\mu\nu}$,
$\delta R_{\mu\nu\lambda\sigma}$, $\delta R_{\mu\nu}$, and $\delta
R$ read,
\begin{align}
\label{Q4} \delta\Gamma^\kappa_{\mu\nu}
=&\frac{1}{2}g^{\kappa\lambda}\left( \nabla_\mu \delta
g_{\nu\lambda} + \nabla_\nu \delta g_{\mu\lambda} - \nabla_\lambda
\delta g_{\mu\nu} \right)\, ,\nn \delta
R_{\mu\nu\lambda\sigma}=&\frac{1}{2}\left[\nabla_\lambda
\nabla_\nu \delta g_{\sigma\mu}
 - \nabla_\lambda \nabla_\mu \delta g_{\sigma\nu}
 - \nabla_\sigma \nabla_\nu \delta g_{\lambda\mu}
 + \nabla_\sigma \nabla_\mu \delta g_{\lambda\nu}
+ \delta g_{\mu\rho} R^\rho_{\ \nu\lambda\sigma}
 - \delta g_{\nu\rho} R^\rho_{\ \mu\lambda\sigma} \right] \, ,\nn
\delta R_{\mu\nu} =& \frac{1}{2}\left[\nabla_\mu\nabla^\rho \delta
g_{\nu\rho} + \nabla_\nu \nabla^\rho \delta g_{\mu\rho} - \Box
\delta g_{\mu\nu}
 - \nabla_\mu \nabla_\nu \left(g^{\rho\lambda}\delta g_{\rho\lambda}\right)
 - 2R^{\lambda\ \rho}_{\ \nu\ \mu}\delta g_{\lambda\rho}
+ R^\rho_{\ \mu}\delta g_{\rho\nu} + R^\rho_{\ \nu}\delta
g_{\rho\mu} \right]\, ,\nn \delta R =& -\delta g_{\mu\nu}
R^{\mu\nu} + \nabla^\mu \nabla^\nu \delta g_{\mu\nu}
 - \Box \left(g^{\mu\nu}\delta g_{\mu\nu}\right)\, .
\end{align}
Accordingly the variation of the Gauss-Bonnet scalar
$\delta\mathcal{G}$ reads,
\begin{align}
\label{FRGBg5} \delta\mathcal{G}=& 2 R \left( -\delta g_{\mu\nu}
R^{\mu\nu} + \nabla^\mu \nabla^\nu \delta g_{\mu\nu}
 - \nabla^2 \left(g^{\mu\nu}\delta g_{\mu\nu}\right) \right)
+ 8 R^{\rho\sigma} R^{\mu\ \nu}_{\ \rho\ \sigma}\delta g_{\mu\nu}
 - 4 \left( R^{\rho\nu} \nabla_\rho\nabla^\mu
+ R^{\rho\mu} \nabla_\rho\nabla^\nu\right) \delta g_{\mu\nu} \nn
& + 4 R^{\mu\nu} \nabla^2 \delta g_{\mu\nu} + 4 R^{\rho\sigma}
\nabla_\rho \nabla_\sigma \left( g^{\mu\nu} \delta g_{\mu\nu}
\right)
 - 2 R^{\mu\rho\sigma\tau} R^\nu_{\ \rho\sigma\tau} \delta g_{\mu\nu}
 - 4 R^{\rho\mu\sigma\nu} \nabla_\rho \nabla_\sigma \delta g_{\mu\nu} \, .
\end{align}
Then for the $f\left( \mathcal{G} \right)$ gravity model
(\ref{GB1b}), by varying the action with respect to the metric
tensor $g_{\mu\nu}$, we obtain the following equations of motion,
\begin{align}
\label{GB4b} 0=& \frac{1}{2\kappa^2}\left(- R^{\mu\nu} +
\frac{1}{2} g^{\mu\nu} R\right) + T_\mathrm{matter}^{\mu\nu} +
\frac{1}{2}g^{\mu\nu} f(\mathcal{G}) + \left( - 2 R R^{\mu\nu} + 8
R^{\rho\sigma} R^{\mu\ \nu}_{\ \rho\ \sigma}
 - 2 R^{\mu\rho\sigma\tau} R^\nu_{\ \rho\sigma\tau}
\right) f' \left( \mathcal{G} \right) \nn & + 2 \left( \nabla^\mu
\nabla^\nu - g^{\mu\nu} \Box \right) \left( R f' \left(
\mathcal{G} \right) \right)
 - 4 \nabla^\mu \nabla_\rho
\left( R^{\rho\nu} f' \left( \mathcal{G} \right) \right)
 - 4 \nabla^\nu \nabla_\rho \left( R^{\rho\mu} f' \left( \mathcal{G} \right) \right)
+ 4 \Box \left( R^{\mu\nu} f' \left( \mathcal{G} \right) \right)
\nn & + 4 g^{\mu\nu} \nabla_\rho \nabla_\sigma \left(
R^{\rho\sigma} f' \left( \mathcal{G} \right) \right) - 4
\nabla_\rho \nabla_\sigma \left( R^{\rho\mu\sigma \nu} f' \left(
\mathcal{G} \right) \right) \, .
\end{align}
By using the Bianchi identities,
\begin{align}
\label{GB5} & \nabla^\rho R_{\rho\tau\mu\nu}= \nabla_\mu
R_{\nu\tau}
 - \nabla_\nu R_{\mu\tau}\, ,\quad
\nabla^\rho R_{\rho\mu} = \frac{1}{2} \nabla_\mu R\, , \nn &
\nabla_\rho \nabla_\sigma R^{\mu\rho\nu\sigma} = \Box R^{\mu\nu}
- \frac{1}{2}\nabla^\mu \nabla^\nu R + R^{\mu\rho\nu\sigma}
R_{\rho\sigma}
 - R^\mu_{\ \rho} R^{\nu\rho} \, , \quad
\nabla_\rho \nabla_\sigma R^{\rho\sigma} = \frac{1}{2} \Box R \, ,
\end{align}
we can rewrite Eq.~(\ref{GB4b}) as follows,
\begin{align}
\label{GB4c} 0= & \frac{1}{2\kappa^2}\left(- R^{\mu\nu} +
\frac{1}{2}g^{\mu\nu} R\right) + T_\mathrm{matter}^{\mu\nu} +
\frac{1}{2}g^{\mu\nu} f(\mathcal{G}) + \left( - 2 R R^{\mu\nu} - 2
R^{\mu\rho\sigma\tau} R^\nu_{\ \rho\sigma\tau} + 4 R^\mu_{\ \rho}
R^{\nu\rho} + 4 R^{\rho\sigma} R^{\mu\ \nu}_{\ \rho\ \sigma}
\right) f' \left( \mathcal{G} \right) \nn & + 2 R \nabla^\mu
\nabla^\nu f' \left( \mathcal{G} \right)
 - 2 g^{\mu\nu} R \Box f' \left( \mathcal{G} \right)
 - 4 R^{\rho\nu} \nabla^\mu \nabla_\rho f' \left( \mathcal{G} \right)
 - 4 R^{\rho\mu} \nabla^\nu \nabla_\rho f' \left( \mathcal{G} \right) \nn
& + 4 R^{\mu\nu} \Box f' \left( \mathcal{G} \right) + 4 g^{\mu\nu}
R^{\rho\sigma} \nabla_\rho \nabla_\sigma f' \left( \mathcal{G}
\right)
 - 4 R^{\rho\mu\sigma \nu} \nabla_\rho \nabla_\sigma f' \left( \mathcal{G} \right)
\, .
\end{align}
Also in four dimensions, we have the following identity,
\begin{equation}
\label{identity} 0= \frac{1}{2}g^{\mu\nu} \mathcal{G}
 - 2 R R^{\mu\nu} - 2 R^{\mu\rho\sigma\tau} R^\nu_{\ \rho\sigma\tau}
+ 4 R^\mu_{\ \rho} R^{\nu\rho} + 4 R^{\rho\sigma} R^{\mu\ \nu}_{\
\rho\ \sigma} \, .
\end{equation}
Then Eq.~(\ref{GB4c}) takes the following form,
\begin{align}
\label{fRGB6} 0 = & \frac{1}{2\kappa^2}\left(- R^{\mu\nu} +
\frac{1}{2}g^{\mu\nu} R\right) + T_\mathrm{matter}^{\mu\nu} +
\frac{1}{2}g^{\mu\nu} \left( f(\mathcal{G}) - \mathcal{G} f'
\left( \mathcal{G} \right) \right) + 2 R \nabla^\mu \nabla^\nu f'
\left( \mathcal{G} \right)
 - 2 g^{\mu\nu} R \Box f' \left( \mathcal{G} \right) \nn
& - 4 R^{\rho\nu} \nabla^\mu \nabla_\rho f' \left( \mathcal{G}
\right)
 - 4 R^{\rho\mu} \nabla^\nu \nabla_\rho f' \left( \mathcal{G} \right)
+ 4 R^{\mu\nu} \Box f' \left( \mathcal{G} \right) + 4 g^{\mu\nu}
R^{\rho\sigma} \nabla_\rho \nabla_\sigma f' \left( \mathcal{G}
\right)
 - 4 R^{\rho\mu\sigma \nu} \nabla_\rho \nabla_\sigma f' \left( \mathcal{G} \right)
\, .
\end{align}
We now rewrite Eq.~(\ref{fRGB6}) in the following form,
\begin{align}
\label{fRGB7} 0 = & \frac{1}{2\kappa^2}\left(- R_{\mu\nu} +
\frac{1}{2}g_{\mu\nu} R\right) + \frac{1}{2} T_{\mathrm{matter}\,
\mu\nu} + \frac{1}{2}g_{\mu\nu} \left( f(\mathcal{G}) -
\mathcal{G} f' \left( \mathcal{G} \right) \right) + D_{\mu\nu}^{\
\ \tau\eta} \nabla_\tau \nabla_\eta f' \left( \mathcal{G} \right)
\, , \nn D_{\mu\nu}^{\ \ \tau\eta} \equiv& \left( \delta_\mu^{\
\tau} \delta_\nu^{\ \eta} + \delta_\nu^{\ \tau} \delta_\mu^{\
\eta} - 2 g_{\mu\nu} g^{\tau\eta} \right) R + \left( - 4
g^{\rho\tau} \delta_\mu^{\ \eta} \delta_\nu^{\ \sigma}
 - 4 g^{\rho\tau} \delta_\nu^{\ \eta} \delta_\mu^{\ \sigma}
+ 4 g_{\mu\nu} g^{\rho\tau} g^{\sigma\eta} \right) R_{\rho\sigma}
\nn & + 4 R_{\mu\nu} g^{\tau\eta}
 - 2 R_{\rho\mu\sigma \nu} \left( g^{\rho\tau} g^{\sigma\eta}
+ g^{\rho\eta} g^{\sigma\tau} \right) \, .
\end{align}
Having in mind that,
\begin{equation}
\label{DDD1} g^{\mu\nu} D_{\mu\nu}^{\ \ \tau\eta} = 4 \left( -
\frac{1}{2} g^{\tau\eta} R + R^{\tau\eta} \right) \, ,
\end{equation}
we find in component form,
\begin{align}
\label{fRGB8}
D_{00}^{\ \ 00} =& 2 R - 2 g_{00} g^{00} R - 8 R^0_{\ 0} + 4 g_{00} R^{00}
+ 4 g^{00} R_{00} - 4 R^{0\ 0}_{\ 0\ 0}\, , \nn
D_{ij}^{\ \ 00} =& 4 g_{ij} R^{00} - 4 R^{0\ 0}_{\ i\ j} - 2 g_{ij} g^{00} R + 4 R_{ij} g^{00} \, .
\end{align}
If we choose the gauge in which $g_{0i}=0$, then the quantity
$D_{00}^{\ \ 00}$ vanishes but $D_{ij}^{\ \ 00}$ does not vanish
in general. This indicates that Eq.~(\ref{fRGB6}) includes the
fourth derivative of metric with respect to the cosmic time
coordinate and therefore ghost modes might appear. We may see the
existence of ghost modes explicitly, by considering perturbations.
Let a solution of (\ref{fRGB6}) be $g_{\mu\nu}=g^{(0)}_{\mu\nu}$
and we denote the curvatures and connections given by
$g^{(0)}_{\mu\nu}$ by using the indexes ``$(0)$''. Then in order
to investigate if any ghost could exist, we may consider the
variation of (\ref{fRGB6}) around the solution $g^{(0)}_{\mu\nu}$
as follows $g_{\mu\nu}=g^{(0)}_{\mu\nu} + \delta g_{\mu\nu}$. For
the variation of $\delta g_{\mu\nu}$, we may impose the following
gauge condition,
\begin{equation}
\label{fRGB9}
0 = \nabla^\mu \delta g_{\mu\nu} \, .
\end{equation}
Then Eq.~(\ref{FRGBg5}) reduces to,
\begin{align}
\label{FRGBg9}
\delta\mathcal{G}=& 2 R \left( -\delta g_{\mu\nu} R^{\mu\nu}
 - \nabla^2 \left(g^{\mu\nu}\delta g_{\mu\nu}\right) \right)
+ 8 R^{\rho\sigma} R^{\mu\ \nu}_{\ \rho\ \sigma}\delta g_{\mu\nu}
+ 4 R^{\mu\nu} \nabla^2 \delta g_{\mu\nu}
+ 4 R^{\rho\sigma} \nabla_\rho \nabla_\sigma
\left( g^{\mu\nu} \delta g_{\mu\nu} \right) \nn
& - 2 R^{\mu\rho\sigma\tau} R^\nu_{\ \rho\sigma\tau} \delta g_{\mu\nu}
 - 4 R^{\rho\mu\sigma\nu} \nabla_\rho \nabla_\sigma \delta g_{\mu\nu} \, .
\end{align}
Even if we impose the condition $\delta g^\mu_{\ \mu}=0$,
Eq.~(\ref{FRGBg9}) has the following form,
\begin{equation}
\label{FRGBg10}
\delta\mathcal{G}= - 2 R R^{\mu\nu} \delta g_{\mu\nu}
+ 8 R^{\rho\sigma} R^{\mu\ \nu}_{\ \rho\ \sigma}\delta g_{\mu\nu}
+ 4 R^{\mu\nu} \nabla^2 \delta g_{\mu\nu}
 - 2 R^{\mu\rho\sigma\tau} R^\nu_{\ \rho\sigma\tau} \delta g_{\mu\nu}
 - 4 R^{\rho\mu\sigma\nu} \nabla_\rho \nabla_\sigma \delta g_{\mu\nu} \, ,
\end{equation}
which also contains the second derivative of the metric
$g_{\mu\nu}$ with respect the cosmic time coordinate. Under the
perturbation $g_{\mu\nu}=g^{(0)}_{\mu\nu} + \delta g_{\mu\nu}$,
the term $D_{\mu\nu}^{\ \ \tau\eta} \nabla_\tau \nabla_\eta f'
\left( \mathcal{G} \right)$ takes the following form,
\begin{equation}
\label{FRGBg11}
D_{\mu\nu}^{\ \ \tau\eta} \nabla_\tau \nabla_\eta f' \left( \mathcal{G} \right)
\to D_{\mu\nu}^{\ \ \tau\eta} \nabla_\tau \nabla_\eta f' \left( \mathcal{G}^{(0)} \right)
+ D_{\mu\nu}^{\ \ \tau\eta} \nabla_\tau \nabla_\eta
\left( f'' \left( \mathcal{G}^{(0)} \right) \delta \mathcal{G} \right) + \cdots \, ,
\end{equation}
which contains the fourth derivative of the metric $g_{\mu\nu}$
with respect to the cosmic time coordinate, and therefore the
perturbed equation (\ref{fRGB7}) may have a ghost mode. Note that
in Eq.~(\ref{FRGBg11}), the ``$\cdots$'' expresses the terms
occurring from the variation of $D_{\mu\nu}^{\ \ \tau\eta}
\nabla_\tau \nabla_\eta$. The propagating mode is a scalar
expressed by the Gauss-Bonnet invariant as it is clear from
Eq.~(\ref{fRGB7}). Having presented explicitly how a ghost mode
may occur in $f(\mathcal{G})$ gravity, we now demonstrate how the
ghost modes may be eliminated or avoided in this theory. This is
the subject of the next subsection.

\subsection{Development of a Ghost-free $f(\mathcal{G})$ Gravity}

In this subsection, we consider how we can avoid the ghost in
$f(\mathcal{G})$ gravity. To this end, we rewrite the action of
Eq.~(\ref{GB1b}) by introducing an auxiliary field $\chi$ as
follows,
\begin{equation}
\label{FRGBg12}
S=\int d^4x\sqrt{-g} \left(\frac{1}{2\kappa^2}R
+ h\left( \chi \right) \mathcal{G} - V\left( \chi \right)
+ \mathcal{L}_\mathrm{matter}\right)\, .
\end{equation}
Then by varying the action (\ref{FRGBg12}) with respect to the
auxiliary field $\chi$, we obtain the following equation,
\begin{equation}
\label{FRGBg13}
0 = h'\left( \chi \right) \mathcal{G} - V'\left( \chi \right) \, ,
\end{equation}
which can be solved with respect to $\chi$ as a function of the
Gauss-Bonnet invariant $\mathcal{G}$ as follows, $\chi = \chi
\left( \mathcal{G} \right)$. Then by substituting the obtained
expression into Eq.~(\ref{FRGBg13}), we reobtain the action of
Eq.~(\ref{GB1b}) with $f \left( \mathcal{G} \right)$ being equal to,
\begin{equation}
\label{FRGBg14}
f \left( \mathcal{G} \right) = h \left( \chi \left( \mathcal{G} \right) \right) \mathcal{G}
 - V \left( \chi \left( \mathcal{G} \right) \right) \, .
\end{equation}
On the other hand, by varying the action (\ref{FRGBg13}) with
respect to the metric tensor, we obtain,
\begin{equation}
\label{FRGBg15}
0 = \frac{1}{2\kappa^2}\left(- R_{\mu\nu}
+ \frac{1}{2}g_{\mu\nu} R\right) + \frac{1}{2} T_{\mathrm{matter}\, \mu\nu}
 - \frac{1}{2}g_{\mu\nu} V \left( \chi \right)
+ D_{\mu\nu}^{\ \ \tau\eta} \nabla_\tau \nabla_\eta h \left( \chi
\right) \, ,
\end{equation}
with $D_{\mu\nu}^{\ \ \tau\eta}$ being defined in
Eq.~(\ref{fRGB7}). Since $\chi$ can be given by a function of the
Gauss-Bonnet invariant $\mathcal{G}$, Eq.~(\ref{FRGBg15}) is the
fourth order differential equation for the metric, which may
actually generate the ghost modes. Eq.~(\ref{FRGBg15}) indicates
that the propagating scalar mode is quantified in terms of $\chi$.
Then in order to make the scalar mode not to be ghost, we may add
a canonical kinetic term of $\chi$ in the action (\ref{FRGBg12})
as in the model of Eq.~(\ref{FRGBg16})~\cite{Nojiri:2005vv}, where
we have chosen the mass dimension of $\chi$ to be unity. Then
instead of Eqs.~(\ref{FRGBg13}) and (\ref{FRGBg15}), we obtain,
\begin{align}
\label{FRGBg17}
0 =& \Box \chi + h'\left( \chi \right) \mathcal{G} - V'\left( \chi \right) \, , \\
\label{FRGBg18}
0 =& \frac{1}{2\kappa^2}\left(- R_{\mu\nu}
+ \frac{1}{2}g_{\mu\nu} R\right) + \frac{1}{2} T_{\mathrm{matter}\, \mu\nu}
+ \frac{1}{2} \partial_\mu \chi \partial_\nu \chi
 - \frac{1}{2}g_{\mu\nu} \left( \frac{1}{2} \partial_\rho \chi \partial^\rho \chi
+ V \left( \chi \right) \right)
+ D_{\mu\nu}^{\ \ \tau\eta} \nabla_\tau \nabla_\eta h \left( \chi \right) \, .
\end{align}
Since the equations derived above do not contain higher than
second order derivatives, if we impose initial conditions for the
following quantities $g_{\mu\nu}$, $\dot g_{\mu\nu}$, $\chi$, and
$\dot\chi$ on a spatial hypersurface of constant cosmic time, the
evolution of $g_{\mu\nu}$ and $\chi$ is uniquely determined, and
as it is clear from Eq.~(\ref{FRGBg18}), these could not be
ghosts. In the model of Eq.~(\ref{FRGBg16}), we have introduced a
new dynamical degree of freedom, namely $\chi$, but if we like to
reduce the dynamical degrees of freedom, we may impose a
constraint as in the mimetic gravity case~\cite{Chamseddine:2013kea,
Nojiri:2014zqa,Dutta:2017fjw}, by
introducing the Lagrange multiplier field $\lambda$, as follows,
\begin{equation}
\label{FRGBg19}
S=\int d^4x\sqrt{-g} \left(\frac{1}{2\kappa^2}R
+ \lambda \left( \frac{1}{2} \partial_\mu \chi \partial^\mu \chi + \frac{\mu^4}{2} \right)
 - \frac{1}{2} \partial_\mu \chi \partial^\mu \chi
+ h\left( \chi \right) \mathcal{G} - V\left( \chi \right) +
\mathcal{L}_\mathrm{matter}\right)\, ,
\end{equation}
where $\mu$ is a constant with mass-dimension one. Then, by
varying the above action (\ref{FRGBg19}) with respect to
$\lambda$, we obtain the constraint,
\begin{equation}
\label{FRGBg20}
0=\frac{1}{2} \partial_\mu \chi \partial^\mu \chi + \frac{\mu^4}{2} \, .
\end{equation}
Then due to the fact that the kinetic term becomes a constant, the
kinetic term in the action of Eq.~(\ref{FRGBg19}) can be absorbed
into the redefinition of the scalar potential $V\left( \chi
\right)$ as follows,
\begin{equation}
\label{FRGBg21}
\tilde V \left(\chi\right) \equiv \frac{1}{2} \partial_\mu \chi \partial^\mu \chi
+ V \left( \chi \right) = - \frac{\mu^4}{2} + V \left( \chi \right) \, ,
\end{equation}
and we can rewrite the action of Eq.~(\ref{FRGBg19}) as follows,
\begin{equation}
\label{FRGBg22}
S=\int d^4x\sqrt{-g} \left(\frac{1}{2\kappa^2}R
+ \lambda \left( \frac{1}{2} \partial_\mu \chi \partial^\mu \chi + \frac{\mu^4}{2} \right)
+ h\left( \chi \right) \mathcal{G} - \tilde V\left( \chi \right)
+ \mathcal{L}_\mathrm{matter}\right)\, .
\end{equation}
For the model of Eq.~(\ref{FRGBg22}), in addition to
Eq.~(\ref{FRGBg20}), we have the following two equations of motion,
\begin{align}
\label{FRGBg23}
0 =& - \frac{1}{\sqrt{-g}} \partial_\mu \left( \lambda g^{\mu\nu}\sqrt{-g}
\partial_\nu \chi \right)
+ h'\left( \chi \right) \mathcal{G} - {\tilde V}'\left( \chi \right) \, , \\
\label{FRGBg24}
0 =& \frac{1}{2\kappa^2}\left(- R_{\mu\nu}
+ \frac{1}{2}g_{\mu\nu} R\right) + \frac{1}{2} T_{\mathrm{matter}\, \mu\nu}
 - \frac{1}{2} \lambda \partial_\mu \chi \partial_\nu \chi
 - \frac{1}{2}g_{\mu\nu} \tilde V \left( \chi \right)
+ D_{\mu\nu}^{\ \ \tau\eta} \nabla_\tau \nabla_\eta h \left( \chi
\right)\, ,
\end{align}
where we have also used Eq.~(\ref{FRGBg20}). By multiplying
Eq.~(\ref{FRGBg24}) with $g^{\mu\nu}$, we obtain,
\begin{equation}
\label{FRGBg24A} 0 = \frac{R}{2\kappa^2} + \frac{1}{2}
T_\mathrm{matter} + \frac{\mu^4}{2} \lambda - 2 \tilde V \left(
\chi \right) - 4 \left( - R^{\tau\eta} + \frac{1}{2} g^{\tau\eta}
R \right) \nabla_\tau \nabla_\eta h \left( \chi \right) \, ,
\end{equation}
where we used Eq.~(\ref{FRGBg20}) and $T_\mathrm{matter} \equiv
g^{\mu\nu} T_{\mathrm{matter}\, \mu\nu}$. Eq.~(\ref{FRGBg24A}) can
be solved with respect to the Lagrange multiplier field $\lambda$,
and the result is,
\begin{equation}
\label{FRGBg24AB}
\lambda = - \frac{2}{\mu^4} \left( \frac{R}{2\kappa^2} + \frac{1}{2} T_\mathrm{matter}
 - 2 \tilde V \left( \chi \right) - 4 \left( - R^{\tau\eta} + \frac{1}{2} g^{\tau\eta} R \right)
\nabla_\tau \nabla_\eta h \left( \chi \right) \right) \, .
\end{equation}
We expect that the model (\ref{FRGBg22}) could not contain a ghost
mode. And actually by using perturbations of the metric, we now
show explicitly that indeed the model (\ref{FRGBg22}) is ghost
free. Let the general solutions of Eqs.~(\ref{FRGBg20}),
(\ref{FRGBg23}), and (\ref{FRGBg24}) be $g^{(0)}_{\mu\nu}$,
$\chi^{(0)}$, and $\lambda^{(0)}$ and we consider the following
perturbation,
\begin{equation}
\label{FRGBg25}
g_{\mu\nu}=g^{(0)}_{\mu\nu}+\delta g_{\mu\nu}\, , \quad
\chi = \chi^{(0)} + \delta \chi\, , \quad
\lambda = \lambda^{(0)} + \delta \lambda \, .
\end{equation}
Then Eqs.~(\ref{FRGBg20}), (\ref{FRGBg23}), and (\ref{FRGBg24})
can be written,
\begin{align}
\label{FRGBg26}
0 = & \partial^\mu \chi^{(0)} \partial_\mu \delta \chi
 - \delta g_{\mu\nu} \partial^\mu \chi^{(0)} \partial^\nu \chi^{(0)} \, , \\
\label{FRGBg27}
0 =& \frac{g^{(0)\, \rho\sigma} \delta g_{\rho\sigma}}{2\sqrt{-g^{(0)}}}
\partial_\mu \left( \lambda^{(0)} g^{(0)\, \mu\nu}\sqrt{-g^{(0)}}
\partial_\nu \chi^{(0)} \right)
 - \frac{1}{\sqrt{-g^{(0)}}}
\partial_\mu \left( \delta \lambda g^{(0)\, \mu\nu}\sqrt{-g^{(0)}}
\partial_\nu \chi^{(0)} \right) \nn
& + \frac{1}{\sqrt{-g^{(0)}}}
\partial_\mu \left( \lambda^{(0)} g^{(0)\, \mu\rho} \delta g_{\rho\sigma} g^{(0)\, \sigma\nu}
\sqrt{-g^{(0)}} \partial_\nu \chi^{(0)} \right)
 - \frac{1}{2\sqrt{-g^{(0)}}}
\partial_\mu \left( \lambda^{(0)} g^{(0)\, \mu\nu}
g^{(0)\, \rho\sigma} \delta g_{\rho\sigma} \sqrt{-g^{(0)}}
\partial_\nu \chi^{(0)} \right) \nn
& - \frac{1}{\sqrt{-g^{(0)}}}
\partial_\mu \left( \lambda^{(0)} g^{(0)\, \mu\nu}\sqrt{-g^{(0)}}
\partial_\nu \delta \chi \right)
+ h''\left( \chi^{(0)} \right) \delta \chi \mathcal{G}^{(0)}
 - {\tilde V}''\left( \chi^{(0)} \right)\delta \chi \nn
& + h'\left( \chi^{(0)} \right) \left( 2 R^{(0)} \left( -\delta g_{\mu\nu} R^{(0)\, \mu\nu}
+ \nabla^{(0)\, \mu} \nabla^{(0)\, \nu} \delta g_{\mu\nu}
 - \Box^{(0)} \left(g^{(0)\,\mu\nu}\delta g_{\mu\nu}\right) \right)
+ 8 R^{(0)\, \rho\sigma} R^{(0)\, \mu\ \nu}_{\ \ \ \ \ \rho\ \sigma}\delta g_{\mu\nu}
\right. \nn
& - 4 \left( R^{(0)\, \rho\nu} \nabla^{(0)}_\rho\nabla^{(0)\, \mu}
+ R^{(0)\, \rho\mu} \nabla^{(0)}_\rho\nabla^{(0)\, \nu}\right) \delta g_{\mu\nu}
+ 4 R^{(0)\, \mu\nu} \Box^{(0)} \delta g_{\mu\nu}
+ 4 R^{(0)\, \rho\sigma} \nabla^{(0)}_\rho \nabla^{(0)}_\sigma
\left( g^{(0)\, \mu\nu} \delta g_{\mu\nu} \right) \nn
& \left. - 2 R^{(0)\, \mu\rho\sigma\tau} R^{(0)\, \nu}_{\ \ \ \ \ \rho\sigma\tau}
\delta g_{\mu\nu}
 - 4 R^{(0)\, \rho\mu\sigma\nu} \nabla^{(0)}_\rho
\nabla^{(0)}_\sigma \delta g_{\mu\nu} \right) \, , \\
\label{FRGBg28}
0 =& \frac{1}{2\kappa^2}\left(- \frac{1}{2}\left(\nabla^{(0)}_\mu
\nabla^{(0)\, \rho} \delta g_{\nu\rho}
+ \nabla^{(0)}_\nu \nabla^{(0)\, \rho} \delta g_{\mu\rho} - \Box^{(0)} \delta g_{\mu\nu}
 - \nabla^{(0)}_\mu \nabla^{(0)}_\nu \left(g^{(0)\, \rho\lambda}\delta g_{\rho\lambda}\right)
\right. \right. \nn
& \left. - 2R^{(0)\, \lambda\ \rho}_{\ \ \ \ \ \nu\ \mu}\delta g_{\lambda\rho}
+ R^{(0)\, \rho}_{\ \ \ \ \ \mu}\delta g_{\rho\nu}
+ R^{(0)\, \rho}_{\ \ \ \ \ \nu}\delta g_{\rho\mu} \right) \nn
& \left. + \frac{1}{2} R^{(0)} \delta g_{\mu\nu}
+ \frac{1}{2}g^{(0)}_{\mu\nu} \left( -\delta g_{\rho\sigma} R^{(0)\, \rho\sigma}
+ \nabla^{(0)\, \rho} \nabla^{(0)\, \sigma} \delta g_{\rho\sigma}
 - \Box^{(0)} \left(g^{(0)\, \rho\sigma}\delta g_{\rho\sigma}\right) \right) \right)
+ \frac{1}{2} \delta T_{\mathrm{matter}\, \mu\nu} \nn
& - \frac{1}{2} \delta \lambda \partial_\mu \chi^{(0)} \partial_\nu \chi^{(0)}
- \frac{1}{2} \lambda^{(0)} \partial_\mu \delta \chi \partial_\nu \chi^{(0)}
- \frac{1}{2} \lambda^{(0)} \partial_\mu \chi^{(0)} \partial_\nu \delta \chi
 - \frac{1}{2}\delta g_{\mu\nu} \tilde V \left( \chi^{(0)} \right)
 - \frac{1}{2}g^{(0)}_{\mu\nu} {\tilde V}' \left( \chi^{(0)} \right) \delta\chi \nn
& + \left\{ - 2 \left( \delta g_{\mu\nu} g^{(0)\, \tau\eta}
 - g^{(0)}_{\mu\nu} g^{(0)\, \tau\zeta} \delta g_{\zeta\xi} g^{(0)\, \xi\eta} \right) R^{(0)}
\right. \nn
& + \left( \delta_\mu^{\ \tau} \delta_\nu^{\ \eta}
+ \delta_\nu^{\ \tau} \delta_\mu^{\ \eta} - 2 g^{(0)}_{\mu\nu} g^{(0)\, \tau\eta} \right)
\left( -\delta g_{\zeta\xi} R^{(0)\, \zeta\xi} + \nabla^{(0)\, \zeta}
\nabla^{(0)\, \xi} \delta g_{\zeta\xi}
 - \Box^{(0)} \left(g^{(0)\, \zeta\xi}\delta g_{\zeta\xi}\right) \right) \nn
& + 4 \left( g^{(0)\, \rho\zeta} \delta g_{\zeta\xi} g^{(0)\, \xi \tau}
\delta_\mu^{\ \eta} \delta_\nu^{\ \sigma}
+ g^{(0)\, \rho\zeta} \delta g_{\zeta\xi} g^{(0)\, \xi\tau}
\delta_\nu^{\ \eta} \delta_\mu^{\ \sigma}
+ \delta g_{\mu\nu} g^{(0)\, \rho\tau} g^{(0)\, \sigma\eta} \right. \nn
& \left. + g^{(0)}_{\mu\nu} g^{(0)\, \rho\zeta} \delta g_{\zeta\xi}
g^{(0)\, \xi\tau} g^{(0)\, \sigma\eta}
+ g^{(0)}_{\mu\nu} g^{(0)\, \rho\tau} g^{(0)\, \sigma\zeta}
\delta g_{\zeta\xi} g^{(0)\, \xi\eta} \right)
R^{(0)}_{\rho\sigma} \nn
& + 2 \left( - g^{(0)\, \rho\tau} \delta_\mu^{\ \eta} \delta_\nu^{\ \sigma}
 - g^{(0)\, \rho\tau} \delta_\nu^{\ \eta} \delta_\mu^{\ \sigma}
+ g^{(0)}_{\mu\nu} g^{(0)\, \rho\tau} g^{(0)\, \sigma\eta} \right)
\left(\nabla^{(0)}_\rho\nabla^{(0)\, \xi} \delta g_{\sigma\xi}
+ \nabla^{(0)}_\sigma \nabla^{(0)\, \xi} \delta g_{\rho\xi} \right. \nn
& \left. - \Box^{(0)} \delta g_{\rho\sigma}
 - \nabla^{(0)}_\rho \nabla^{(0)}_\sigma
\left(g^{(0)\, \zeta\xi}\delta g_{\zeta\xi}\right)
 - 2R^{(0)\, \zeta\ \xi}_{\ \ \ \ \ \sigma\ \rho}\delta g_{\zeta\xi}
+ R^{(0)\, \xi}_{\ \ \ \ \ \rho}\delta g_{\xi\sigma}
+ R^{(0)\, \xi}_{\ \ \ \ \ \sigma}\delta g_{\xi\rho} \right) \nn
& - 4 R^{(0)}_{\mu\nu} g^{(0)\, \tau\zeta} \delta g_{\zeta\xi} g^{(0)\, \xi\eta}
+ 4 g^{(0)\, \tau\eta} \left(\nabla^{(0)}_\mu\nabla^{(0)\, \xi} \delta g_{\nu\xi}
+ \nabla^{(0)}_\nu \nabla^{(0)\, \xi} \delta g_{\mu\xi} \right. \nn
& \left. - \Box^{(0)} \delta g_{\mu\nu} - \nabla^{(0)}_\mu \nabla^{(0)}_\nu
\left(g^{(0)\, \zeta\xi}\delta g_{\zeta\xi}\right)
 - 2R^{(0)\, \zeta\ \xi}_{\ \ \ \ \ \nu\ \mu}\delta g_{\zeta\xi}
+ R^{(0)\, \xi}_{\ \ \ \ \ \mu}\delta g_{\xi\nu}
+ R^{(0)\, \xi}_{\ \ \ \ \ \nu}\delta g_{\xi\mu} \right) \nn
& -2 \left( \nabla^{(0)}_\sigma \nabla^{(0)}_\mu \delta g_{\nu\rho}
 - \nabla^{(0)}_\sigma \nabla^{(0)}_\rho \delta g_{\nu\mu}
 - \nabla^{(0)}_\nu \nabla^{(0)}_\mu \delta g_{\sigma\rho}
 + \nabla^{(0)}_\nu \nabla^{(0)}_\rho \delta g_{\sigma\mu}
+ \delta g_{\rho\xi} R^{(0)\, \xi}_{\ \ \ \ \ \mu\sigma\nu}
 - \delta g_{\mu\xi} R^{(0)\, \xi}_{\ \ \ \ \ \rho\sigma\nu} \right)
g^{(0)\, \rho\tau} g^{(0)\, \sigma\eta} \nn
& \left. - 4 R^{(0)}_{\rho\mu\sigma \nu} \left(
g^{(0)\, \rho\zeta} \delta g_{\zeta\xi} g^{(0)\, \xi\tau} g^{(0)\, \sigma\eta}
+ g^{(0)\, \rho\tau} g^{(0)\, \sigma\zeta} \delta g_{\zeta\xi} g^{(0)\, \xi\eta} \right) \right\}
\nabla^{(0)}_\tau \nabla^{(0)}_\eta h \left( \chi^{(0)} \right) \nn
& - \frac{1}{2} D_{\mu\nu}^{(0) \ \tau\eta}
g^{(0)\, \zeta\xi}\left(
\nabla^{(0)}_\tau \delta g_{\eta\xi}
+ \nabla^{(0)}_\eta \delta g_{\tau\xi} - \nabla^{(0)}_\xi \delta g_{\tau\eta}
\right) \partial_\zeta h \left( \chi^{(0)} \right)
+ D_{\mu\nu}^{(0) \ \tau\eta} \nabla^{(0)}_\tau \nabla^{(0)}_\eta
\left( h' \left( \chi^{(0)} \right) \delta \chi \right) \, .
\end{align}
On the other hand, Eq.~(\ref{FRGBg24AB}) yields,
\begin{align}
\label{FRGBg29}
\delta \lambda =& - \frac{2}{\mu^4} \left( \frac{1}{2\kappa^2}
\left( -\delta g_{\rho\sigma} R^{(0)\, \rho\sigma}
+ \nabla^{(0)\, \rho} \nabla^{(0)\, \sigma} \delta g_{\rho\sigma}
 - \Box^{(0)} \left(g^{(0)\, \rho\sigma}\delta g_{\rho\sigma}\right) \right)
+ \frac{1}{2} \delta T_\mathrm{matter} \right. \nn
& - 2 \tilde V' \left( \chi^{(0)} \right) \delta \chi
 - 4 \left(- \frac{1}{2}\left(g^{(0)\, \eta\xi} \nabla^{(0)\, \tau}
\nabla^{(0)\, \rho} \delta g_{\xi\rho}
+ g^{(0)\, \tau\xi} \nabla^{(0)\, \eta} \nabla^{(0)\, \rho} \delta g_{\xi\rho}
\right. \right. \nn
& \left. - g^{(0)\, \tau\xi} g^{(0)\, \eta\zeta} \Box^{(0)} \delta g_{\xi\zeta}
 - \nabla^{(0)\, \tau} \nabla^{(0)\, \eta} \left(g^{(0)\, \rho\lambda}
\delta g_{\rho\lambda}\right)
 - 2R^{(0)\, \lambda\eta\rho\tau}\delta g_{\lambda\rho}
+ R^{(0)\, \rho\tau} g^{(0)\, \eta\xi} \delta g_{\rho\xi}
+ R^{(0)\, \rho\eta} g^{(0)\, \tau\xi} \delta g_{\rho\xi} \right) \nn
& \left. + \frac{1}{2} R^{(0)} g^{(0)\, \tau\zeta} g^{\eta\xi} \delta g_{\zeta\xi}
+ \frac{1}{2}g^{(0)\, \tau\eta} \left( -\delta g_{\rho\sigma} R^{(0)\, \rho\sigma}
+ \nabla^{(0)\, \rho} \nabla^{(0)\, \sigma} \delta g_{\rho\sigma}
 - \Box^{(0)} \left(g^{(0)\, \rho\sigma}\delta g_{\rho\sigma}\right) \right) \right)
\nabla^{(0)}_\tau \nabla^{(0)}_\eta h \left( \chi^{(0)} \right) \nn
& - 4 \left( g^{(0)\, \tau\zeta} \delta g_{\zeta\xi} g^{(0)\, \xi\alpha} g^{(0)\, \eta\beta}
+ g^{(0)\, \eta\zeta} \delta g_{\zeta\xi} g^{(0)\, \xi\alpha} g^{(0)\, \tau\beta} \right)
\left( - R^{(0)}_{\alpha\beta} + \frac{1}{2} g^{(0)}_{\alpha\beta} R^{(0)} \right)
\nabla^{(0)}_\tau \nabla^{(0)}_\eta h \left( \chi^{(0)} \right) \nn
& - 4 \left( - R^{(0)\, \tau\eta} + \frac{1}{2} g^{(0)\, \tau\eta} R^{(0)} \right)
\left( - \frac{1}{2} D_{\mu\nu}^{(0) \ \tau\eta}
g^{(0)\, \zeta\xi}\left(
\nabla^{(0)}_\tau \delta g_{\eta\xi}
+ \nabla^{(0)}_\eta \delta g_{\tau\xi} - \nabla^{(0)}_\xi \delta g_{\tau\eta}
\right) \partial_\eta h \left( \chi^{(0)} \right) \right. \nn
& \left. \left. + D_{\mu\nu}^{(0) \ \tau\eta} \nabla^{(0)}_\tau \nabla^{(0)}_\eta
\left( h' \left( \chi^{(0)} \right) \delta \chi \right) \right) \right) \, .
\end{align}
By substituting Eq.~(\ref{FRGBg29}) in Eq.~(\ref{FRGBg28}), we
may eliminate $\delta\lambda$. The obtained equation contains
first and second derivatives of $\delta g_{\mu\nu}$ and $\chi$,
especially the first and second derivatives with respect to the
cosmic time $t$. We can choose $\chi^{(0)}$ to be,
\begin{equation}
\label{frgdS4B}
\chi^{(0)} = \mu^2 t \, .
\end{equation}
Then Eq.~(\ref{FRGBg26}) takes the following form,
\begin{equation}
\label{FRGBg30} 0 = \delta \dot\chi - \mu^2 \delta g_{tt} \, ,
\end{equation}
and we also have $\delta \ddot\chi = \mu^2 \delta {\dot g}_{tt}$.
Then we can further eliminate the variation terms $\delta
\dot\chi$ and $\delta \ddot\chi$, and the obtained equation
contains the first and second derivatives of $\delta g_{\mu\nu}$
with respect to the cosmic time $t$, but does not include the
first and second derivative terms $\delta \chi$ again with respect
to time $t$. Then by providing the initial conditions for $\delta
g_{\mu\nu}$, $\delta {\dot g}_{\mu\nu}$, and $\chi$ on a spatial
hypersurface, we can determine the time evolution of $\delta
g_{\mu\nu}$ uniquely up to the gauge invariance corresponding to
the general covariance of the model, and the corresponding
constraints. This indicates that the number of the physical
degrees of freedom is only two. Eq.~(\ref{FRGBg30}) also indicates
that $\chi$ is not dynamical and the time evolution of $\chi$ is
given by Eq.~(\ref{FRGBg30}). Therefore, no additional degrees of
freedom occur, compared to the standard Einstein-Hilbert gravity,
and in effect, no ghost modes actually occur in the theory. Having
demonstrated that the modified $f(\mathcal{G})$ gravity theory can
be rendered ghost-free, let us consider several examples of
cosmological evolutions which can be realized in the context of
this theory. This is the subject of the next subsection.


\subsection{Boundary terms of Ghost-free $f\left( \mathcal{G} \right)$ Gravity}

In the present paper, our main interest for deriving the
ghost-free equations of motion, is on cosmological applications,
so the boundary terms should be of no particular interest. We
shall come to this issue soon, however it is worthy to discuss the
differences that certain boundary terms would bring along, in the
case that one is interested in working on spacetimes with
boundaries. In this case, for spacetimes $M$ with boundary
$\partial M$, the variation of the action (\ref{FRGBg22}) induces
the following terms on the boundary,
\begin{align}
\label{bdry1}
\delta S_\mathrm{boundary} = \int_{\partial M} d^3 x \sqrt{ - q} & \left[ \frac{1}{2\kappa^2}
\left\{ n^\mu \nabla^\nu \delta g_{\mu\nu}
 - n^\rho \nabla_\rho \left( g^{\mu\nu} \delta g_{\mu\nu} \right)
\right\} \right. \nn
& + h(\chi) \left\{
R n^\mu \nabla^\nu \delta g_{\mu\nu} + R n^\nu \nabla^\mu \delta g_{\mu\nu}
 - \left( \nabla^\mu R \right) n^\nu \delta g_{\mu\nu}
 - \left( \nabla^\nu R \right) n^\mu \delta g_{\mu\nu} \right. \nn
& - 4 n_\rho \left( R^{\rho\nu} \nabla^\mu \delta g_{\mu\nu}
+ R^{\rho\mu} \nabla^\nu \delta g_{\mu\nu} \right)
+ 4 \left( n^\mu \left( \nabla_\rho R^{\rho\nu} \right)
+ n^\nu \left( \nabla_\rho R^{\rho\mu} \right) \right) \delta g_{\mu\nu}  \nn
& + 4 n_\rho R^{\mu\nu} \nabla^\rho \delta g_{\mu\nu}
 - 4 n^\rho \left( \nabla_\rho R^{\mu\nu} \right) \delta g_{\mu\nu}
+ 4 n_R^{\rho\sigma} \nabla_\sigma \left( g^{\mu\nu} \delta g_{\mu\nu} \right)
 - 4 n_\sigma \left( \nabla_\rho R^{\rho\sigma} \right)
g^{\mu\nu} \delta g_{\mu\nu} \nn
& \left. - 4 n_\rho R^{\rho\mu\sigma\nu} \nabla_\sigma \delta g_{\mu\nu}
+ 4 n_\sigma \left(\nabla_\rho R^{\rho\mu\sigma\nu} \right) \delta g_{\mu\nu}
\right\} \nn
& + h'(\chi) \left\{ - R \left( \nabla^\mu \chi \right) n^\nu \delta g_{\mu\nu}
 - R \left( \nabla^\nu \chi \right) n^\mu \delta g_{\mu\nu}
+ 4 \left( n^\mu R^{\rho\nu} \left( \nabla_\rho \chi \right)
+ n^\nu R^{\rho\mu} \left( \nabla_\rho \chi \right) \right) \delta g_{\mu\nu} \right. \nn
& \left. - 4 n^\rho R^{\mu\nu} \left( \nabla_\rho \chi \right) \delta g_{\mu\nu}
 - 4 n_\sigma R^{\rho\sigma} \left( \nabla_\rho \chi \right)
g^{\mu\nu} \delta g_{\mu\nu} + 4 n_\sigma R^{\rho\mu\sigma\nu}
\left(\nabla_\rho \chi \right) \delta g_{\mu\nu} \right\} \nn &
\left. - \lambda n^\mu \partial_\mu \chi \right] \, ,
\end{align}
where $n_\mu$ is a unit vector ($n_\mu n_\mu =1$ if $n_\mu$ is
space-like and $n_\mu n_\mu =-1$ if $n_\mu$ is time-like) which is
perpendicular and outward to the boundary and $l_{\mu\nu} \equiv
g_{\mu\nu} - n_\mu n_\nu$ is the induced metric on the boundary
and $l$ is the determinant of $l_{\mu\nu}$. In order for the
variational principle to be well-defined, we need to require
$\delta S_\mathrm{boundary}=0$, which cannot be realized because
$\delta S_\mathrm{boundary}$ includes both of $\delta g_{\mu\nu}$
without derivative and $\nabla_\sigma \delta g_{\mu\nu}$. In order
to avoid this problem, we can add Gibbons-Hawking boundary terms
\cite{Gibbons:1976ue},
\begin{equation}
\label{bdry2}
S_\mathrm{GH} = \frac{1}{\kappa^2} \int_{\partial M} d^3
\sqrt{-l} l^{\mu\nu} \nabla_\mu n_\nu \, ,
\end{equation}
for the part of the Einstein-Hilbert term, which is proportional
to $\frac{1}{\kappa^2}$ or Myers-like boundary terms
\cite{Myers:1987yn},
\begin{align}
\label{bdry3}
S_\mathrm{M} =& 2 \int_{\partial M} d^3 x \sqrt{-l} h(\chi) \left\{
\frac{1}{3} \left( 2 K K_{\mu\rho} K^{\rho\nu} + K_{\rho\sigma} K^{\rho\sigma} K
 - 2 K_{\mu\rho} K^{\rho\sigma} K_\sigma^{\ \mu} - K^3 \right)
 - G_l^{\mu\nu} K_{\mu\nu} \right\} \, , \nn
K_{\mu\nu} \equiv& l^\rho_{\ \mu} l^\sigma_{\ \nu} \nabla_\rho n_\sigma\, ,
\end{align}
for the terms proportional to $h(\phi)$ but not to $h'(\phi)$. For
some recent useful applications of Gibbons-Hawking like terms in
Euclidean gravity, see \cite{Cai:2008ht,Ro:2016kyu}. Particularly,
the terms proportional to $h'(\phi)$ and $\lambda$ give some
boundary conditions for the scalar fields $\chi$ and $\lambda$,
which could be, for example,
\begin{equation}
\label{sbdry4}
h'(\chi)=0\, , \quad \lambda=0 \, ,
\end{equation}
which could correspond to the boundary conditions chosen in Refs.
\cite{Cai:2008ht,Ro:2016kyu}.

However in the most cases of a homogeneous and isotropic metric in
cosmology, the most characteristic type of metric chosen is a FRW
metric, with or without spatial curvature. In the flat FRW case,
the topological spaces not excluded from the data up-to-date is
the three-torus which is flat in 4-dimensional spacetime and the
infinite 3-Euclidean plane, in which case no boundaries occur,
unless some strong finite-time singularity occurs in the future.
In that case, the singularities which lead to geodesics
incompleteness, like the Big Rip, may lead eventually to having
certain forms of boundaries on the spacelike hypersurface on which
the singularities occur, but this effect is hard to quantify with
Gibbons-Hawking terms because the ending of a future timelike
geodesic is highly non-trivial to define from a mathematical point
of view, so no induced metric can be defined on it, and actually
closed timelike curves can occur and at the same time be absorbed
in the same notion of the future singularity. Some useful
treatment of these issues can be found in \cite{Tipler:1977eb}. So
we refrain to further discuss the boundary terms issue, which is
however useful for non-cosmological applications.


\section{FRW Cosmology in Ghost-free $f\left( \mathcal{G} \right)$ Gravity}

In this section, we consider the cosmology produced by the
ghost-free $f\left( \mathcal{G} \right)$ gravity model of
Eq.~(\ref{FRGBg22}). Especially we show that it is possible to realize
any cosmological era of the Universe, by using the model under
consideration. We will particularly try to realize the late and
early-time acceleration eras.

\subsection{A Reconstruction Technique for Model Building}

Let us firstly demonstrate how the equations of motion of the
model (\ref{FRGBg22}) become in the case the metric is a flat
Friedman-Robertson-Walker metric (FRW) with line element,
\begin{equation}
\label{FRWmetric} ds^2 = - dt^2 + a(t)^2 \sum_{i=1,2,3} \left(
dx^i \right)^2 \, .
\end{equation}
For this metric, we have,
\begin{align}
\label{E2}
& \Gamma^t_{ij}=a^2 H \delta_{ij}\, ,\quad \Gamma^i_{jt}=\Gamma^i_{tj}=H\delta^i_{\ j}\, ,
\quad \Gamma^i_{jk}=\tilde \Gamma^i_{jk}\, ,\quad
R_{itjt}=-\left(\dot H + H^2\right)a^2\delta_{ij}\, ,\quad
R_{ijkl}= a^4 H^2 \left(\delta_{ik} \delta_{lj}
 - \delta_{il} \delta_{kj}\right)\, ,\nn
& R_{tt}=-3\left(\dot H + H^2\right)\, ,\quad R_{ij}=a^2
\left(\dot H + 3H^2\right)\delta_{ij}\, ,\quad R= 6\dot H + 12
H^2\, , \quad \mbox{other components}=0\, , \nn & \mathcal{G} = 24
H^2 \left( \dot H + H^2 \right) \, ,
\end{align}
where $H \equiv \frac{\dot a}{a}$. We also assume that $\lambda$
and $\chi$ depend solely on the cosmic time $t$, that is,
$\lambda=\lambda(t)$ and $\chi=\chi(t)$. We also assume
$T_{\mathrm{matter}\, \mu\nu} =0$ just for simplicity. Then a
solution of Eq.~(\ref{FRGBg20}) is given below,
\begin{equation}
\label{frgdS4}
\chi = \mu^2 t \, .
\end{equation}
In effect, the $(t,t)$ component and $(i,j)$ component of
(\ref{FRGBg24}) yield,
\begin{align}
\label{FRGFRW1}
0 = & - \frac{3H^2}{2\kappa^2} - \frac{\mu^4 \lambda}{2}
+ \frac{1}{2} \tilde V \left( \mu^2 t \right) - 12 \mu^2 H^3 h' \left( \mu^2 t \right) \, ,
\\
\label{FRGFRW2}
0 = & \frac{1}{2\kappa^2} \left( 2 \dot H + 3 H^2 \right)
 - \frac{1}{2} \tilde V \left( \mu^2 t \right)
+ 4 \mu^4 H^2 h'' \left( \mu^2 t \right)
+ 8 \mu^2 \left( \dot H + H^2 \right) H h' \left( \mu^2 t \right) \, .
\end{align}
On the other hand, Eq,~(\ref{FRGBg23}) gives,
\begin{equation}
\label{FRGFRW3}
0 = \mu^2 \dot\lambda + 3 \mu^2 H \lambda
+ 24 H^2 \left( \dot H + H^2 \right) h'\left( \mu^2 t \right)
 - {\tilde V}'\left( \mu^2 t \right) \, , \\
\end{equation}
Eq.~(\ref{FRGFRW1}) can be solved with respect to $\lambda$ as follows,
\begin{equation}
\label{FRGFRW4}
\lambda = - \frac{3 H^2}{\mu^4 \kappa^2} + \frac{1}{\mu^4} \tilde V \left( \mu^2 t \right)
- \frac{24}{\mu^2} H^3 h' \left( \mu^2 t \right) \, .
\end{equation}
Then by substituting Eq.~(\ref{FRGFRW4}) into Eq. (\ref{FRGFRW3}),
we reobtain Eq.~(\ref{FRGFRW2}). On the other hand,
Eq.~(\ref{FRGFRW2}) can be solved with respect to $\tilde V \left(
\mu^2 t \right)$ as follows,
\begin{equation}
\label{FRGFRW7}
\tilde V \left( \mu^2 t \right) = \frac{1}{\kappa^2} \left( 2 \dot H + 3 H^2 \right)
+ 8 \mu^4 H^2 h'' \left( \mu^2 t \right)
+ 16 \mu^2 \left( \dot H + H^2 \right) H h' \left( \mu^2 t \right) \, ,
\end{equation}
which tells that for arbitrary $h(\chi)$, if the potential $\tilde
V \left( \chi \right)$ is assumed to be equal to,
\begin{equation}
\label{FRGFRW8}
\tilde V \left( \chi \right) = \left[ \frac{1}{\kappa^2} \left( 2 \dot H + 3 H^2 \right)
+ 8 \mu^4 H^2 h'' \left( \mu^2 t \right)
+ 16 \mu^2 \left( \dot H + H^2 \right) H h' \left( \mu^2 t \right)
\right]_{t=\frac{\chi}{\mu^2}}\, ,
\end{equation}
then an arbitrary cosmological evolution of the Universe with
Hubble rate $H=H(t)$ can be realized. By combining
Eqs.~(\ref{FRGFRW4}) and (\ref{FRGFRW7}), we also obtain,
\begin{equation}
\label{FRGFRW4B}
\lambda = \frac{2 \dot H}{\mu^4 \kappa^2}
+ 8 H^2 h'' \left( \mu^2 t \right)
+ \frac{8}{\mu^2} \left( 2 \dot H - H^2 \right) H h' \left( \mu^2 t \right) \, .
\end{equation}
Basically the above procedure is a reconstruction method for the
model (\ref{FRGBg22}) and by using this method it is possible to
realize an arbitrarily given cosmological evolution. In the next
subsection we shall use this reconstruction method.

\subsection{Early and Late-time Accelerating Universe Cosmologies with Ghost-free $f(\mathcal{G})$ Gravity}

In this subsection, we consider some examples of models which
describe an accelerating Universe. As a first example, we consider
a de Sitter space-time realization, in which case the Hubble rate
$H$ is a constant $H=H_0$. Then by using Eq.~(\ref{FRGFRW8}), for
an arbitrarily chosen function $h(\chi)$, the corresponding scalar
potential is given by,
\begin{equation}
\label{FRGFRW9}
\tilde V \left( \chi \right) = \frac{3 H_0^2}{\kappa^2}
+ 8 \mu^4 H_0^2 h'' \left( \chi \right)
+ 16 \mu^2 H_0^3 h' \left( \chi \right) \, .
\end{equation}
Eq.~(\ref{FRGFRW4B}) also indicates how the Lagrange multiplier
$\lambda$ in this model behaves, and it is equal to,
\begin{equation}
\label{FRGFRW10}
\lambda \left( t \right) = 8 H_0^2 h'' \left( \mu^2 t \right)
 - \frac{8}{\mu^2} H_0^3 h' \left( \mu^2 t \right) \, .
\end{equation}
Then by appropriately choosing the functional form of $h\left(
\chi \right)$, we can obtain several different ghost-free $f\left(
\mathcal{G} \right)$ models which can realize a de Sitter
evolution. Next we consider the model which mimics the
$\Lambda$CDM model, in which case the Hubble rate $H$ is given by,
\begin{equation}
\label{FRGFRW11}
H = H_0 \coth \left( \frac{3}{2} H_0 t \right) \, .
\end{equation}
At late times, that is in the limit $t\to + \infty$, $H$ in
Eq.~(\ref{FRGFRW11}) behaves as follows,
\begin{equation}
\label{FRGFRW12}
H \to H_0 \, ,
\end{equation}
which corresponds to an asymptotic de Sitter spacetime. On the
other hand, at early times, which era is reached in the limit
$t\to 0$, the Hubble rate behaves as follows,
\begin{equation}
\label{FRGFRW11B}
H \to \frac{3}{2t} \, ,
\end{equation}
which corresponds to a matter or dust dominated Universe. Then by
using Eq.~(\ref{FRGFRW8}), we find,
\begin{equation}
\label{FRGFRW12B}
\tilde V \left( \chi \right) = \frac{3H_0^2}{\kappa^2}
+ 8 \mu^4 H_0^2 \coth^2 \left( \frac{3H_0}{2\mu^2} \chi \right)
h'' \left( \chi \right)
+ 16 \mu^2 H_0^3 \left( 1 - \frac{1}{2\sinh^2\left( \frac{3H_0}{2\mu^2} \chi \right)} \right)
\coth \left( \frac{3H_0}{2\mu^2} \chi \right) h' \left( \chi \right) \, ,
\end{equation}
and from Eq.~(\ref{FRGFRW4B}) we can determine the functional form
of the Lagrange multiplier $\lambda$, which is,
\begin{equation}
\label{FRGFRW13}
\lambda = \frac{3 H_0^2}{\mu^4 \kappa^2 \sinh^2\left( \frac{3}{2} H_0 t \right)}
+ 8 H_0^2 \coth^2 \left( \frac{3}{2} H_0 t \right) h'' \left( \mu^2 t \right)
 - \frac{8H_0^3}{\mu^2} \left( 1 +
\frac{4}{\sinh^2\left( \frac{3}{2} H_0 t \right)} \right)
\coth \left( \frac{3}{2} H_0 t \right) h' \left( \mu^2 t \right) \, .
\end{equation}
The model of Eq.~(\ref{FRGFRW11}), which is generated in the
context of ghost-free $f(\mathcal{G})$ gravity by the scalar
potential of Eq.~(\ref{FRGFRW12B}), realizes the $\Lambda$CDM
model without introducing any dark matter perfect fluid.
Therefore, the model incorporates the cosmological constant part,
corresponding to an equation of state (EoS) parameter being equal
to $w=-1$, and also incorporates the cold dark matter (CDM) part,
corresponding to and EoS parameter exactly equal to $w=0$. Thus we
have succeeded to realize the present accelerating expansion of
the Universe by using the ghost-free $f\left( \mathcal{G} \right)$
gravity model. Notably, the cosmological evolution
(\ref{FRGFRW11}) can be realized in the context of the ghost-free
$f(\mathcal{G})$ by using a function $h\left( \chi \right)$ and an
arbitrary parameter $\mu^2$. In the case of the standard
Einstein-Hilbert gravity, the FRW equations have the following
form,
\begin{equation}
\label{EoS1} \frac{3}{\kappa^2} H^2 = \rho_\mathrm{total}\, ,
\quad - \frac{1}{\kappa^2} \left( 2 \dot H + 3 H^2 \right) =
p_\mathrm{total} \, ,
\end{equation}
where $\rho_\mathrm{total}$ and $p_\mathrm{total}$ are the total
energy density and the total pressure. In effect, the total
equation of state (EoS) parameter $w_\mathrm{total}$ defined by
$w_\mathrm{total}=\frac{p_\mathrm{total}}{\rho_\mathrm{total}}$ is
equal to,
\begin{equation}
\label{EoS2}
w_\mathrm{total} = - 1 - \frac{2 \dot H}{3 H^2}\, .
\end{equation}
We should note that the effective total EoS parameter
$w_\mathrm{total}$ includes the contributions of all the fluid
components of the Universe like the dark energy, dark matter, and
so on. The Planck 2018 results~\cite{Aghanim:2018eyx}, constrain
the Hubble constant, which is the present value of the Hubble
rate, as follows $H_\mathrm{present} = \left(67.4\pm 0.5\right)\,
\mathrm{km}\, \mathrm{s}^{-1}\, \mathrm{Mpc}^{-1}$. Also the
matter density parameter is constrained as $\Omega_\mathrm{m} =
0.315\pm 0.007$ and finally, the dark energy EoS parameter is
constrained as $w_0 = -1.03 \pm 0.03$ although $w_0$ is different
from $w_\mathrm{total}^\mathrm{eff}$. Since $p_\mathrm{total} =
\left( 1 - \Omega_\mathrm{m} \right) w_0 \rho_\mathrm{total}$, the
Planck 2018 results indicate that,
\begin{equation}
\label{EoS5_0}
w_\mathrm{total} = \left( 1 - \Omega_\mathrm{m} \right) w_0
\sim - 0.705\, .
\end{equation}
Even for a general modified gravity theory, in the case of the
ghost-free $f\left( \mathcal{G} \right)$ gravity case we developed
in this paper, the effective total EoS parameter
$w_\mathrm{total}^\mathrm{eff}$ is defined in Eq.~(\ref{EoS2}),
that is,
\begin{equation}
\label{EoS3}
w_\mathrm{total}^\mathrm{eff} = - 1 - \frac{2 \dot H}{3 H^2}\, .
\end{equation}
Then in the case of the de Sitter space as in the model
(\ref{FRGFRW8}) in this paper, since the Hubble rate is a
constant, $H=H_0$, we find $w_\mathrm{total}^\mathrm{eff} = - 1$.
On the other hand, in the case of the model mimicking the
$\Lambda$CDM model, namely model (\ref{FRGFRW11}), we find,
\begin{equation}
\label{EoS4} w_\mathrm{total}^\mathrm{eff} = - 1 -
\frac{1}{\cosh^2 \left( \frac{3}{2} H_0 t_\mathrm{present}
\right)}\, ,
\end{equation}
where $t_\mathrm{present}$ is the value of the cosmic time today.
In the model (\ref{FRGFRW11}), the dark matter contribution to the
evolution is effectively included. Then the Planck 2018 results
(\ref{EoS5_0}) constrain the parameters of the model
(\ref{FRGFRW11}). Due to the fact that the observed Hubble
constant is $H_\mathrm{present} = \left(67.4\pm 0.5\right)\,
\mathrm{km}\, \mathrm{s}^{-1}\, \mathrm{Mpc}^{-1}$, by using
(\ref{FRGFRW11}) we find,
\begin{equation}
\label{FRGFRW11BB}
H_0 \coth \left( \frac{3}{2} H_0 t_\mathrm{present} \right)
= \left(67.4\pm 0.5\right)\, \mathrm{km}\, \mathrm{s}^{-1}\, \mathrm{Mpc}^{-1} \, .
\end{equation}
On the other hand, combined with Eq.~(\ref{EoS4}), the Planck 2018
results (\ref{EoS5_0}) indicate that,
\begin{equation}
\label{EoS5}
\frac{1}{\cosh^2 \left( \frac{3}{2} H_0 t_\mathrm{present} \right)} \sim 0.294\, .
\end{equation}
Then Eqs.~(\ref{FRGFRW11BB}) and (\ref{EoS5}) actually constrain
the parameters $H_0$ and $t_\mathrm{present}$ of the model, so
these can appropriately be chosen so that the constraints are
satisfied.

In addition, since the $\Lambda$CDM model is still consistent with
any constraint obtained from the observations on the current
expansion of the Universe, the model (\ref{FRGFRW11}) mimicking
the $\Lambda$CDM model should be consistent with the current
observational data. In the future, perhaps some deviations from
the standard $\Lambda$CDM model may be observed. Then by using the
formulation of ghost-free $f\left( \mathcal{G} \right)$ gravity
model which we presented in this paper, we can always construct a
more realistic model than the $\Lambda$CDM model, according to
future observations.

As another model, we shall consider the following cosmological
model with parameters, $\delta$, $H_0$, $H_i$, $t_s$, $\mu$, and $\Lambda$,
\begin{equation}
\label{hubblenew}
H(t)=\delta \e^{H_0-H_i t} \tanh \left( \frac{t_s-t}{\mu} \right)+\Lambda\, ,
\end{equation}
where the parameters $\mu$ and $H_i$ are measured in seconds in
natural units, while the parameter $\delta$ has dimensions
sec$^{-2}$ in natural units. In addition, the parameter $H_0$ is
considered to be dimensionless. The above model has quite
interesting early and late-time phenomenology if the free
parameters are appropriately chosen, since it can qualitatively
describe a quasi-de Sitter cosmological evolution at early times
and an accelerating era of de Sitter form at late times. Indeed,
if the parameter $t_s$ is chosen to be the age of the present
Universe, and also if the parameter $\Lambda$ is chosen to be the
present time cosmological constant, then at early times when $t
\ll t_s$, the first term is approximated as follows,
\begin{equation}
\label{earlytimeapprox}
H(t)\sim \delta \left(\e^{H_0}- \e^{H_0} H_i t \right)-\Lambda\, ,
\end{equation}
due to the fact that at early times,
\begin{equation}
\label{earltanh}
\tanh \left(\frac{t_s-t}{\mu} \right) \sim 1\, .
\end{equation}
Hence, if $H_0$ and $H_i$ are appropriately chosen so that
$\e^{H_0}\, ,H_i\gg \Lambda$, the early-time evolution is a
quasi-de Sitter evolution of the form,
\begin{equation}
\label{quasidesitterearlytimeapprox}
H(t)\sim \delta \left( \e^{H_0}-\e^{H_0} H_i t \right)\, ,
\end{equation}
and the effective EoS parameter is nearly
$w_\mathrm{total}^\mathrm{eff}\sim -1$. Accordingly, at late-times
when $t\sim t_s$, the exponential in Eq.~(\ref{hubblenew}) tends
to zero, and also we have,
\begin{equation}
\label{latetanh}
\tanh \left( \frac{t_s-t}{\mu} \right) \sim 0\, ,
\end{equation}
in effect, the Hubble rate is again approximated by an exact de
Sitter evolution,
\begin{equation}
\label{exactdesitter}
H(t)\sim \Lambda\, .
\end{equation}
The realization of the model (\ref{earlytimeapprox}) in the
context of the ghost free $f(\mathcal{G})$ is possible, if the
scalar potential is equal to,
\begin{align}
\label{scalarpotential}
V(\chi(t))=&\frac{3 \left(\e^{H_0-H_i t} \tanh \left(\frac{t_s-t}{\mu }\right)
+\Lambda \right)^2-\frac{2\e^{H_0-H_i t}
\left(H_i \mu \tanh \left(\frac{t_s-t}{\mu }\right)
+ \mathrm{sech}^2\left(\frac{t_s-t}{\mu }\right)\right)}{\mu}}{\kappa ^2} \nn
& +16 \mu ^2 \left( \e^{H_0-H_i t} \tanh \left(\frac{t_s-t}{\mu }\right)
+\Lambda \right) \nn
& \times h'(\chi ) \left(\left(\e^{H_0-H_i t} \tanh \left(\frac{t_s-t}{\mu }\right)
+\Lambda \right)^2-H_i \e^{H_0-H_i t} \tanh \left(\frac{t_s-t}{\mu }\right)
-\frac{\e^{H_0-H_i t} \mathrm{sech}^2\left(\frac{t_s-t}{\mu }\right)}{\mu }\right) \nn
& +8 \mu ^4 h''(\chi ) \left( \e^{H_0-H_i t} \tanh
\left(\frac{t_s-t}{\mu }\right)+\Lambda \right)^2\, .
\end{align}
and the Lagrange multiplier function $\lambda (\chi (t))$ is
chosen as,
\begin{align}
\label{lagrangemultiplierfunction}
\lambda (\chi (t)) =& \frac{2 \left(-H_i \e^{H_0-H_i t} \tanh \left(\frac{t_s-t}{\mu }\right)
-\frac{\e^{H_0-H_i t} \mathrm{sech}^2\left(\frac{t_s-t}{\mu }\right)}{\mu }\right)}
{\kappa^2 \mu ^4}
+\frac{8 h'(\chi )}{\mu^2} \left( \e^{H_0-H_i t}
\tanh \left(\frac{t_s-t}{\mu }\right)+\Lambda \right) \nn
& \times \left(2 \left(-H_i \e^{H_0-H_i t} \tanh \left(\frac{t_s-t}{\mu}\right)
 -\frac{\e^{H_0-H_i t} \mathrm{sech}^2\left(\frac{t_s-t}{\mu}\right)}{\mu}\right)
 -\left(\e^{H_0-H_i t} \tanh\left(\frac{t_s-t}{\mu }\right)+\Lambda \right)^2\right) \nn
& + 8 h''(\chi ) \left(\e^{H_0-H_i t} \tanh\left(\frac{t_s-t}{\mu }\right)
+\Lambda \right)^2\, .
\end{align}
By appropriately choosing the function $h(\chi )$, one may obtain
different models which can realize the same cosmological evolution
(\ref{earlytimeapprox}), so a rich phenomenology can be obtained.
The scalar potential at early times is much more simplified, since
it takes the form,
\begin{align}
\label{earlyscalarpotential}
& V(\chi (t))\sim \frac{3 \left( \e^{H_0} - \e^{H_0} H_i t\right)^2 -2 \e^{H_0} H_i}
{\kappa ^2} \nn
& +8 \mu ^4 h''(\chi ) \left( \e^{H_0} - \e^{H_0} H_i t\right)^2
+16 \mu ^2 h'(\chi ) \left( \e^{H_0} - \e^{H_0} H_i t\right)
\left(\left( \e^{H_0} - \e^{H_0} H_i t\right)^2 - \e^{H_0} H_i\right)\, ,
\end{align}
while at late-times it is approximated by,
\begin{equation}
\label{latetimesscalarpotential}
V(\chi (t))\sim 8 \Lambda ^2 \mu ^4 h''(\chi )+16 \Lambda ^3 \mu^2 h'(\chi )
+\frac{3 \Lambda ^2}{\kappa ^2}\, .
\end{equation}
The most interesting feature of the ghost-free model can be seen
by looking Eqs.~(\ref{earlyscalarpotential}) and
(\ref{latetimesscalarpotential}), due to the presence of the
function $h(\chi)$ in both equations. This means that by
appropriately choosing the function $h(\chi )$ so that a viable
early-time phenomenology is obtained, this choice will affect the
late-time phenomenology to some extent, not via the late-time
Hubble rate, but certainly through the scalar potential and the
Lagrange multiplier function $\lambda$. Therefore, quite
interesting phenomenologies may be obtained, due to the fact that
during the two eras the EoS parameter is nearly
$w_\mathrm{total}^\mathrm{eff}\sim -1$, hence the potential and
the Lagrange multiplier function may affect other observable
quantities and render the model more compatible with the
observational data. Work is in progress towards this direction.

Before closing this section we should note that other cosmological
evolutions can be realized in the context of the ghost-free
$f(\mathcal{G})$ theory which developed. For example, consider the
symmetric bounce with Hubble rate,
\begin{equation}\label{symmetricbounce}
H(t)= \e^{\alpha t^2}\, ,
\end{equation}
which a well known bounce cosmology~\cite{Brandenberger:2016vhg,deHaro:2015wda}.
The symmetric bounce
has interesting phenomenology, since in the limit $t\to -\infty$,
the EoS parameter is approximately,
$w_\mathrm{total}^\mathrm{eff}\sim -1$, which is a nearly de
Sitter phase. After that and as the bouncing point at $t=0$ is
approached, the Universe experiences quintessential acceleration
which gradually turns to a decelerating expansion, with gradually
negative and positive EoS parameter. Near the bouncing point, the
Universe experiences another nearly de Sitter accelerating era,
and as the cosmic time grows it is followed by a phantom
accelerating era, which eventually tends to a nearly de Sitter
expansion at $t\to \infty$. It is conceivable that the most
interesting part of this bounce cosmology, from a phenomenological
point of view, is the contracting phase. This cosmological
evolution can be realized by the scalar potential,
\begin{equation}\label{scalarbounce}
V(\chi)=\frac{\e^{\alpha t^2} \left(8 \kappa ^2 \mu ^4 \e^{\alpha
t^2} h''(\chi )+16 \kappa ^2 \mu ^2 \e^{\alpha t^2}
\left(\e^{\alpha t^2}+2 \alpha t\right) h'(\chi )+3 \e^{\alpha
t^2}+4 \alpha t\right)}{\kappa ^2}\, ,
\end{equation}
where $\chi=t\mu^2$, and also by the Lagrange multiplier function
$\lambda (\chi)$,
\begin{equation}\label{lambdabounce}
\lambda (\chi)=8 \e^{2 \alpha t^2} h''(\chi )+\frac{8 \e^{\alpha
t^2} \left(4 \alpha t \e^{\alpha t^2} - \e^{2 \alpha t^2}\right)
h'(\chi )}{\mu ^2}+\frac{4 \alpha t \e^{\alpha t^2}}{\kappa ^2
\mu ^4}\, ,
\end{equation}
where in both Eqs.~(\ref{scalarbounce}) and (\ref{lambdabounce}),
the function $h(\chi)$ is arbitrary. Thus in the context of the
formalism we developed, we do not have a single model realizing
the symmetric bounce, but a class of models which can realize this
cosmological evolution. In principle, the choice of the function
$h(\chi)$ can be done in such a way so that the phenomenological
constraints can be satisfied. We do not further discuss this topic
for brevity, but it is conceivable that there is much room for
realizing interesting phenomenologies.


\section{Newton Law in Ghost-free $f(\mathcal{G})$ Gravity}

In this section we shall consider the Newton law in the context of
ghost-free $f(\mathcal{G})$ and we shall investigate how this
becomes in the ghost free theory. Some alternative solutions in
the context of general Gauss-Bonnet theories can be found in Refs.
\cite{Antoniou:2017acq,Lee:2018zym}. In order to consider the
correction to the Newton law, we assume the geometric background
is flat, by considering the limit of $H\to 0$ in the last section.
This is because we like to consider the Newton law at scales much
smaller in comparison to the cosmological scales, which are of the
order $\sim \frac{1}{H}$ in an asymptotically de Sitter spacetime
during the present time era of the Universe. Then
Eq.~(\ref{FRGFRW7}) or Eq.~(\ref{FRGFRW8}) indicate that $\tilde V
\left( \chi \right)=0$ although $h \left( \chi \right)$ can be an
arbitrary function in general. Therefore Eq.~(\ref{FRGFRW4})
suggests that $\lambda = \lambda^{(0)}=0$. We also assume that the
gauge condition (\ref{fRGB9}) holds true. Then by using
Eqs.~(\ref{frgdS4B}), (\ref{FRGBg26}), (\ref{FRGBg27}),
(\ref{FRGBg28}), and (\ref{FRGBg29}), we obtain,
\begin{align}
\label{FRGBg26B}
0 = & - \mu^2 \partial_t \delta \chi - \mu^4 \delta g_{tt} \, , \\
\label{FRGBg27B}
0 =& \mu^2 \partial_t \delta \lambda
 - h'\left( \chi^{(0)} \right) \Box^{(0)} \left(\eta^{\mu\nu}\delta g_{\mu\nu}\right) \, , \\
\label{FRGBg28B}
0 =& - \frac{1}{4\kappa^2}\left( - \Box^{(0)} \delta g_{\mu\nu}
 - \partial_\mu \partial_\nu \left(\eta^{\rho\lambda}\delta g_{\rho\lambda}\right)
+ g^{(0)}_{\mu\nu} \Box^{(0)} \left(\eta^{\rho\sigma}\delta g_{\rho\sigma}\right)
\right) + \frac{1}{2} \delta T_{\mathrm{matter}\, \mu\nu}
 - \frac{1}{2} \mu^4 \delta_{t\mu} \delta_{t\nu} \delta \lambda \nn
& + \mu^4 \left\{ 2 \eta_{\mu\nu}
\Box^{(0)} \left(\eta^{\zeta\xi}\delta g_{\zeta\xi}\right)
+ 2 \left( - \eta^{\rho t} \delta_\mu^{\ t} \delta_\nu^{\ \sigma}
 - \eta^{\rho t} \delta_\nu^{\ t} \delta_\mu^{\ \sigma}
+ \eta_{\mu\nu} \eta^{\rho t} \eta^{\sigma t} \right)
\left( - \Box^{(0)} \delta g_{\rho\sigma}
 - \partial_\rho \partial_\sigma \left(\eta^{\zeta\xi}\delta g_{\zeta\xi}\right) \right) \right. \nn
& \left. - 4 \left( - \Box^{(0)} \delta g_{\mu\nu}
 - \partial_\mu \partial_\nu \left(\eta^{\zeta\xi}\delta g_{\zeta\xi}\right) \right)
 -2 \left( \partial_t \partial_\mu \delta g_{\nu t}
 - \partial_t^2 \delta g_{\nu\mu}
 - \partial_\nu \partial_\mu \delta g_{tt}
 + \partial_\nu \partial_t \delta g_{t\mu} \right) \right\} h'' \left( \chi^{(0)} \right) \, , \\
\label{FRGBg29B}
\delta \lambda =& - \frac{2}{\mu^4} \left( - \frac{1}{2\kappa^2}
\Box^{(0)} \left(\eta^{\rho\sigma}\delta g_{\rho\sigma}\right)
+ \frac{1}{2} \delta T_\mathrm{matter}
 - 2 \mu^4 \left( \Box^{(0)} \delta g_{tt}
+ \partial_t^2 \left(\eta^{\rho\lambda} \delta g_{\rho\lambda}\right)
+ \Box^{(0)} \left(\eta^{\rho\sigma}\delta g_{\rho\sigma}\right) \right)
h'' \left( \chi^{(0)} \right) \right) \, .
\end{align}
By substituting Eq.~(\ref{FRGBg29B}) in (\ref{FRGBg28B}), we
obtain,
\begin{align}
\label{FRGBg28C}
0 =& - \frac{1}{4\kappa^2}\left( - \Box^{(0)} \delta g_{\mu\nu}
 - \partial_\mu \partial_\nu \left(\eta^{\rho\lambda}\delta g_{\rho\lambda}\right)
+ g^{(0)}_{\mu\nu} \Box^{(0)} \left(\eta^{\rho\sigma}\delta g_{\rho\sigma}\right)
\right) + \frac{1}{2} \delta T_{\mathrm{matter}\, \mu\nu} \nn
& + \delta_{t\mu} \delta_{t\nu} \left\{ - \frac{1}{2\kappa^2}
\Box^{(0)} \left(\eta^{\rho\sigma}\delta g_{\rho\sigma}\right)
+ \frac{1}{2} \delta T_\mathrm{matter}
 - 2 \mu^4 \left( \Box^{(0)} \delta g_{tt}
+ \partial_t^2 \left(\eta^{\rho\lambda} \delta g_{\rho\lambda}\right)
+ \Box^{(0)} \left(\eta^{\rho\sigma}\delta g_{\rho\sigma}\right) \right)
h'' \left( \chi^{(0)} \right) \right\} \nn
& + \mu^4 \left\{ 2 \eta_{\mu\nu}
\Box^{(0)} \left(\eta^{\zeta\xi}\delta g_{\zeta\xi}\right)
 + 2 \left( - \eta^{\rho t} \delta_\mu^{\ t} \delta_\nu^{\ \sigma}
 - \eta^{\rho t} \delta_\nu^{\ t} \delta_\mu^{\ \sigma}
+ \eta_{\mu\nu} \eta^{\rho t} \eta^{\sigma t} \right)
\left( - \Box^{(0)} \delta g_{\rho\sigma}
 - \partial_\rho \partial_\sigma \left(\eta^{\zeta\xi}\delta g_{\zeta\xi}\right) \right)
\right. \nn
& \left. - 4 \left( - \Box^{(0)} \delta g_{\mu\nu}
 - \partial_\mu \partial_\nu \left(\eta^{\zeta\xi}\delta g_{\zeta\xi}\right) \right)
 -2 \left( \partial_t \partial_\mu \delta g_{\nu t}
 - \partial_t^2 \delta g_{\nu\mu}
 - \partial_\nu \partial_\mu \delta g_{tt}
 + \partial_\nu \partial_t \delta g_{t\mu} \right) \right\} h'' \left( \chi^{(0)} \right) \, .
\end{align}
We shall consider a static point gravitational source for the
matter at the spatial origin, that is,
\begin{equation}
\label{Nwtn1} \delta T_{\mathrm{matter}\, tt} = M \delta^{(3)}
\left( \bm{x} \right) \, , \quad \mbox{other components of }
\delta T_{\mathrm{matter}\, \mu\nu} = 0 \, ,
\end{equation}
where $\left( \bm{x} \right) = \left( x^i \right)$. In the
following two subsections, we shall investigate how the Newton law
is modified in the context of Lagrange multiplier constrained
Einstein-Hilbert gravity and in the context of ghost-free
$f(\mathcal{G})$ gravity.

\subsection{Newton Law for Lagrange Multiplier Constrained Einstein-Hilbert gravity}

Let us first consider the constrained Einstein-Hilbert gravity
case, in which case $h\left( \chi \right)=0$. Then
Eq.~(\ref{FRGBg28C}) reduces to,
\begin{align}
\label{FRGBg28D}
0 =& - \frac{1}{4\kappa^2}\left\{ - \Box^{(0)} \delta g_{\mu\nu}
 - \partial_\mu \partial_\nu \left(\eta^{\rho\lambda}\delta g_{\rho\lambda}\right)
+ g^{(0)}_{\mu\nu} \Box^{(0)} \left(\eta^{\rho\sigma}\delta g_{\rho\sigma}\right) \right\}
+ \frac{1}{2} \delta T_{\mathrm{matter}\, \mu\nu} \nn
& + \delta_{t\mu} \delta_{t\nu} \left\{ - \frac{1}{2\kappa^2}
\Box^{(0)} \left(\eta^{\rho\sigma}\delta g_{\rho\sigma}\right)
+ \frac{1}{2} \delta T_\mathrm{matter} \right\} \, .
\end{align}
The $(t,t)$, $(i,j)$, and $(t,i)$ components of (\ref{FRGBg28D})
yield,
\begin{align}
\label{FRGBg28tt}& 0 = - \frac{1}{4\kappa^2}\left\{ - \Box^{(0)}
\delta g_{tt}
 - \partial_t^2 \left(\eta^{\rho \lambda}\delta g_{\rho \lambda}\right)
 \Box^{(0)} \left(\eta^{\rho\sigma}\delta
g_{\rho\sigma}\right) \right\} \, , \\ &\label{FRGBg28ij} 0 = -
\frac{1}{4\kappa^2}\left\{ - \Box^{(0)} \delta g_{ij}
 - \partial_i \partial_j \left(\eta^{\rho\lambda}\delta g_{\rho\lambda}\right)
+ \delta_{ij} \Box^{(0)} \left(\eta^{\rho\sigma}\delta
g_{\rho\sigma}\right) \right\} \, , \\ & \label{FRGBg28ti} 0 = -
\frac{1}{4\kappa^2}\left\{ - \Box^{(0)} \delta g_{ti}
 - \partial_t \partial_i\nu \left(\eta^{\rho\lambda}\delta g_{\rho\lambda}\right) \right\}
\, .
\end{align}
and Eq.~(\ref{FRGBg29B}) has the following form,
\begin{equation}
\label{FRGBg29C}
\delta \lambda = - \frac{2}{\mu^4} \left( - \frac{1}{2\kappa^2}
\Box^{(0)} \left(\eta^{\rho\sigma}\delta g_{\rho\sigma}\right)
+ \frac{1}{2} \delta T_\mathrm{matter} \right) \, .
\end{equation}
We now assume that,
\begin{equation}
\label{Nwtn2} \delta g_{tt} = A(r)\, , \quad \delta g_{ij} =
B(r)\delta_{ij} + C(r) x^i x^j \, , \quad \delta g_{ti} =0 \, ,
\end{equation}
where $r= \sqrt{ \sum_{i=1,2,3} \left( x^i \right)^2}$. Then
Eq.~(\ref{FRGBg28ti}) is trivially satisfied and since,
\begin{align}
\label{calculation}
\eta^{\rho\lambda}\delta g_{\rho\lambda} = & - A + 3 B + r^2 C \, , \nn
\triangle{(x^i x^j C(r) )}
=& 2 \delta_{ij} C(r) + \frac{6 x^i x^j}{r} C'(r) + x^i x^j C''(r) \nn
\partial_i \partial_j ( - A + 3 B + r^2 C )
= & \frac{\delta_{ij}}{r} \left( - A' + 3 B' + 2r C + r^2 C' \right)
+ \frac{x^ix^j}{r^3} \left(
A' - r A'' - 3 B' + 3r B'' + 3 r^2 C' + r^3 C'' \right) \nn
\triangle{( - A + 3 B + r^2 C )} = & \frac{1}{r}\left( - 2 A' - r A'' + 6 B' + 3 r B''
+ 6 r C + 6 r^2 C' + r^3 C''' \right)
\end{align}
Eqs.~(\ref{FRGBg28tt}), (\ref{FRGBg28ij}) have the following forms,
\begin{align}
\label{FRGBg28ttB}
0 =& \frac{1}{4\kappa^2}\triangle{\left( 3 B + r^2 C \right)} \, , \\
\label{FRGBg28ijB}
0 =& - \frac{1}{4\kappa^2}\left\{
 - \frac{\delta_{ij}}{r} \left( A' + r A'' - B' - 2 r B'' - 2 r C -5 r^2 C'
 - r^3 C'' \right) \right. \nn
& \left. - \frac{x^ix^j}{r^3} \left( A' - r A'' - 3 B' + 3r B'' + 9 r^2 C' + 2 r^3 C'' \right)
\right\} \, .
\end{align}
In effect, we have,
\begin{align}
\label{cal1}
0 =& 3 B + r^2 C \, , \\
\label{cal2}
0 =& A' + r A'' - B' - 2 r B'' - 2 r C -5 r^2 C' - r^3 C'' \, , \\
\label{cal3}
0 =& A' - r A'' - 3 B' + 3r B'' + 9 r^2 C' + 2 r^3 C'' \, .
\end{align}
By using Eq.~(\ref{cal1}), we can eliminate $B$ from
Eqs.~(\ref{cal2}) and (\ref{cal3}), so we get,
\begin{align}
\label{cal4}
0 =& A' + r A'' - 2 r^2 C' - \frac{r^3}{3} C'' \, , \\
\label{cal5}
0 =& A' - r A'' + 6 r^2 C' + r^2 C'' \, .
\end{align}
By also eliminating $C$ from Eqs.~(\ref{cal4}) and (\ref{cal5}),
we obtain,
\begin{equation}
\label{cal6}
0 = 4 A' + 2 r A'' \, .
\end{equation}
Under the boundary condition that $A\to 0$ when $r\to \infty$, the
solution of Eq.~(\ref{cal6}) is given by,
\begin{equation}
\label{cal6B}
A = \frac{A_0}{r} \, ,
\end{equation}
with a constant $A_0$. Then Eq.~(\ref{cal2}) takes the following
form,
\begin{equation}
\label{cal7}
0 = \frac{A_0}{r^2} - \frac{1}{3r^3} \left( r^6 C' \right)' \, ,
\end{equation}
and a solution of the above equation is,
\begin{equation}
\label{cal8} C = - \frac{A_0}{2r^3}\, .
\end{equation}
In effect, Eq.~(\ref{cal1}) indicates that,
\begin{equation}
\label{cal9} B=\frac{A_0}{6r}\, ,
\end{equation}
where we have assumed that the boundary condition $B,C \to 0$ when
$r\to \infty$ holds true. Eq.~(\ref{FRGBg29C}) also suggests that,
\begin{equation}
\label{cal10}
\delta \lambda = - \frac{2}{\mu^4} \left( - \frac{4\pi A_0}{2 \kappa^2}
\delta^{(3)} \left( \bm{x} \right)
+ \frac{1}{2} M \delta^{(3)} \left( \bm{x} \right) \right)\, .
\end{equation}
If we put $\delta \lambda = 0$, we find,
\begin{equation}
\label{cal11}
A_0 = \frac{\kappa^2 M}{4\pi} \, ,
\end{equation}
which reproduces the standard Newtonian potential
$\phi_\mathrm{Newton}$, that is,
\begin{equation}
\label{cal12} \phi_\mathrm{Newton} \equiv \frac{A}{2} =
\frac{\kappa^2 M}{8\pi} = \frac{GM}{r} \, ,
\end{equation}
where $G = \frac{\kappa^2}{8\pi}$ is the Newton gravitational
constant. We should note, however, that Eq.~(\ref{cal10})
indicates that there is an infinite number of solutions, which do
not always reproduce the standard Newton law if $\delta\lambda\neq
0$. In addition, Eq.~(\ref{FRGBg27B}) indicates that $0
=\partial_t \delta \lambda$ if $h=0$, which corresponds to the
Einstein-Hilbert gravity case. Therefore, if we put $\delta\lambda
=0$ as an initial condition, then the term $\delta\lambda$ always
vanishes, and the model reproduces the standard Newton law.

\subsection{Newton Law in Ghost-free $f\left( \mathcal{G} \right)$ gravity}

Let us now investigate how the Newton law becomes in the context of
the ghost-free $f\left( \mathcal{G} \right)$ gravity model
(\ref{FRGBg22}). First we assume that Eq.~(\ref{Nwtn2}) holds true
in this case too. Then the general solutions of
Eqs.~(\ref{FRGBg26B}) and (\ref{FRGBg27B}) are given by,
\begin{equation}
\label{Nwtn2B} \delta \chi = - \mu^2 t A(r) + c_1 \left( \bm{x}
\right) \, , \quad \delta \lambda = \frac{1}{\mu^2 r^2 } h \left(
\mu^2 t \right) \left( r^2 \left( - A(r) + 3 B(r) + r^2 C(r)
\right)' \right)' + c_2 \left( \bm{x} \right) \, ,
\end{equation}
where $c_1 \left( \bm{x} \right)$ and $c_2 \left( \bm{x} \right)$
appear by integrating with respect to $t$, and these can be
determined by Eq.~(\ref{FRGBg28B}). However by assuming a
spherical symmetry, then $c_1 \left( \bm{x} \right)$ and $c_2
\left( \bm{x} \right)$ should depend on $\bm{x}$ via the radial
coordinate $r$, that is, $c_1 \left( \bm{x} \right)=c_1 \left( r
\right)$ and $c_2 \left( \bm{x} \right)=c_2 \left( r \right)$. On
the other hand, Eq.~(\ref{FRGBg29B}) has the following form,
\begin{equation}
\label{Nwtn3}
\delta \lambda = - \frac{2}{\mu^4} \left\{ - \frac{1}{2\kappa^2}
\left( r^2 \left( - A(r) + 3 B(r) + r^2 C(r) \right)' \right)'
 - \frac{M}{2} \delta^{(3)} \left( \bm{x} \right)
 - \frac{2 \mu^4}{r^2} \left( r^2 \left( 3 B(r) + r^2 C(r) \right)' \right)'
h'' \left( \mu^2 t \right) \right\} \, .
\end{equation}
By comparing $\delta\lambda$ from Eq.~(\ref{Nwtn2B}) with
(\ref{Nwtn3}), for arbitrary $h\left( \chi \right)$, we find
$A(r)=3B(r) + r^2 C(r) =0$ and $c_2 (r) = - \frac{M}{2}
\delta^{(3)} \left( \bm{x} \right)$. If surely $A(r)=0$, the
result is in conflict with the resulting Newton law of the
constrained Einstein gravity case, given in Eq.~(\ref{cal12}).
This indicates that the assumption (\ref{Nwtn2}) is not satisfied
and the correction to the Newton law should be time-dependent,
which could constrain $\mu^2$, $h \left( \chi \right)$, and/or
$\tilde V \left( \chi \right)$, so that the correction could be
consistent with any experiment or observation.
Eq.~(\ref{FRGBg28C}) indicates that the correction to the Newton law
in the case of Einstein-Hilbert gravity is proportional to the
parameter $\mu^4$ and the function $h\left( \chi \right)$, and
therefore if $\mu^4$ or $h\left( \chi \right)$ are small enough,
the constraint for the Newton law is always satisfied. For the
case of the model (\ref{FRGFRW11}) which mimics the $\Lambda$CDM
model, as long as we consider the Newton law at scales much
smaller than the cosmological scales $\sim \frac{1}{H}$ and as
long as $h\left( \chi \right)$ is small, the constraint for the
Newton law is independent on the cosmological constraints. So the
constraints (\ref{FRGFRW11BB}) and (\ref{EoS5}) can be imposed
without restricting $\mu$ and $h\left( \chi \right)$.

\section{Ghost-Free $F\left( R, \mathcal{G} \right)$ gravity}

As a final task we shall demonstrate how to obtain a ghost-free
$F\left( R, \mathcal{G} \right)$ theory of gravity. The vacuum
$F\left( R, \mathcal{G} \right)$ gravity action is,
\begin{equation}
\label{FRGBg1} S = \int d^4 x \sqrt{-g} F\left( R, \mathcal{G}
\right)\, ,
\end{equation}
where $F\left( R, \mathcal{G} \right)$ is a function of the scalar
curvature $R$ and $\mathcal{G}$ stands for the Gauss-Bonnet
invariant given in Eq.~(\ref{GB}). It was claimed that this model
(\ref{FRGBg1}) has ghost instabilities~\cite{DeFelice:2009ak}, so
let us see how ghost degrees of freedom are manifested at the
equations of motion level. By introducing two auxiliary fields
$\Phi$ and $\Theta$, the action of Eq.~(\ref{FRGBg1}) can be
rewritten as follows,
\begin{equation}
\label{FRGBg2}
S = \int d^4 x \sqrt{-g} \left\{ \frac{\Phi R}{2\kappa^2} + \Theta \mathcal{G}
 - V\left( \Phi, \Theta \right) \right\}\, ,
\end{equation}
where we have introduced the gravitational coupling $\kappa$ in
order to make $\Phi$ and $\Theta$ dimensionless. By varying the
action (\ref{FRGBg2}) with respect to $\Phi$ and $\Theta$, we
obtain,
\begin{equation}
\label{FRGBg3}
\frac{R}{2\kappa^2} = \frac{\partial V\left( \Phi, \Theta \right)}{\partial \Phi} \, ,
\quad \mathcal{G} = \frac{\partial V\left( \Phi, \Theta \right)}{\partial \Theta} \, ,
\end{equation}
which can be algebraically solved with respect $\Phi$ and
$\Theta$, that is, $\Phi = \Phi \left( R, \mathcal{G} \right)$ and
$\Theta = \Theta \left( R, \mathcal{G} \right)$. Then by
substituting the obtained expressions for $\Phi = \Phi \left( R,
\mathcal{G} \right)$ and $\Theta = \Theta \left( R, \mathcal{G}
\right)$ in Eq.~(\ref{FRGBg2}), we obtain the action
(\ref{FRGBg1}) with,
\begin{equation}
\label{FRGBg4}
F\left( R, \mathcal{G} \right) = \Phi \left( R, \mathcal{G} \right) R
+ \Theta \left( R, \mathcal{G} \right) \mathcal{G}
 - V\left( \Phi \left( R, \mathcal{G} \right),
\Theta \left( R, \mathcal{G} \right) \right) \, .
\end{equation}
In order to investigate the properties of the action
(\ref{FRGBg2}), we work in the Einstein frame, so under a
conformal transformation of the form $g_{\mu\nu}\to \e^\phi
g_{\mu\nu}$, the curvatures are transformed as
follows~\cite{Maeda:1988ab,Nojiri:2003ft},
\begin{align}
\label{E4}
R_{\zeta\mu\rho\nu} \to& \left\{R_{\zeta\mu\rho\nu}
 - \frac{1}{2}\left(g_{\zeta\rho}\nabla_\nu \nabla_\mu \phi
+ g_{\mu\nu}\nabla_\rho \nabla_\zeta \phi
 - g_{\mu\rho} \nabla_\nu \nabla_\zeta \phi
 - g_{\zeta\nu} \nabla_\rho \nabla_\mu \phi \right) \right. \nn
& + \frac{1}{4}\left(g_{\zeta\rho}\partial_\nu \phi \partial_\mu \phi
+ g_{\mu\nu} \partial_\rho \phi \partial_\zeta \phi
 - g_{\mu\rho} \partial_\nu \phi \partial_\zeta \phi
 - g_{\zeta\nu} \partial_\rho \phi \partial_\mu \phi \right) \nn
& - \frac{1}{4}\left(g_{\zeta\rho} g_{\mu\nu}
 - g_{\zeta\nu} g_{\mu\rho}\right)\partial^\sigma \phi
\partial_\sigma \phi \Bigr\}\, ,\nn
R_{\mu\nu} \to& R_{\mu\nu} - \frac{1}{2}\left( 2\nabla_\mu \nabla_\nu \phi
+ g_{\mu\nu} \Box \phi\right)
+ \frac{1}{2}\partial_\mu \phi \partial_\nu \phi - \frac{1}{2} g_{\mu\nu}\partial^\sigma \phi
\partial_\sigma \phi\, ,\nn
R \to& \left(R - 3 \Box \phi - \frac{3}{2}\partial^\sigma \phi \partial_\sigma \phi
\right)\e^{-\phi}\, .
\end{align}
Therefore the Gauss-Bonnet invariant $\mathcal{G}$ in
Eq.~(\ref{GB}) is transformed in the following way,
\begin{equation}
\label{fG6}
\mathcal{G} \to \e^{-2\phi} \left[ \mathcal{G} + \nabla_\mu \left\{
4 \left( R^{\mu\nu} - \frac{1}{2} g^{\mu\nu} R \right) \partial_\nu \phi
+ 2 \left( \partial^\mu \phi \Box \phi
 - \left( \nabla_\nu \nabla^\mu \phi\right) \partial^\nu \phi \right)
+ \partial_\nu \phi \partial^\nu \phi \partial^\mu \phi \right\} \right] \, .
\end{equation}
Then by writing $\Phi=\e^{-\phi}$, the action of
Eq.~(\ref{FRGBg2}) can be rewritten by taking into account the
conformal transformation $g_{\mu\nu}\to \e^\phi g_{\mu\nu}$ as
follows,
\begin{align}
\label{FRGBg4b}
S =& \int d^4 x \sqrt{-g} \left\{ \frac{1}{2\kappa^2} \left( R
 - \frac{3}{2}\partial^\sigma \phi \partial_\sigma \phi \right) \right. \nn
& \left. + \Theta \mathcal{G} - \partial_\mu \Theta \left\{
4 \left( R^{\mu\nu} - \frac{1}{2} g^{\mu\nu} R \right) \partial_\nu \phi
+ 2 \left( \partial^\mu \phi \Box \phi
 - \left( \nabla_\nu \nabla^\mu \phi\right) \partial^\nu \phi \right)
+ \partial_\nu \phi \partial^\nu \phi \partial^\mu \phi \right\}
 - \e^{2\phi} V\left( \e^{-\phi}, \Theta \right) \right\}
\, .
\end{align}
This action (\ref{FRGBg4b}) may have ghost degrees of freedom due
to the existence of $\Theta$. As in the last section, we might
eliminate the ghost degrees of freedom by writing $\Theta$ as
$\Theta = \e^\theta$ and add a constraint to the action
(\ref{FRGBg4b}) by using the Lagrange multiplier field $\lambda$,
in the following way,
\begin{align}
\label{FRGBg4c}
S =& \int d^4 x \sqrt{-g} \left\{ \frac{1}{2\kappa^2} \left( R
 - \frac{3}{2}\partial^\sigma \phi \partial_\sigma \phi
 - \lambda\left( \partial_\mu \theta \partial^\mu \theta + \mu^2 \right)
\right) \right. \nn
& \left. + \e^\theta \mathcal{G} - \e^\theta\partial_\mu \theta \left\{
4 \left( R^{\mu\nu} - \frac{1}{2} g^{\mu\nu} R \right) \partial_\nu \phi
+ 2 \left( \partial^\mu \phi \Box \phi
 - \left( \nabla_\nu \nabla^\mu \phi\right) \partial^\nu \phi \right)
+ \partial_\nu \phi \partial^\nu \phi \partial^\mu \phi \right\}
 - \e^{2\phi} V\left( \e^{-\phi}, \e^\theta \right) \right\}
\, .
\end{align}
As in the previous section, the scalar fields $\theta$ and
$\lambda$ are not dynamical degrees of freedom and the dynamical
degrees of freedom are actually the metric and the scalar field
$\phi$, as in the standard $F(R)$ gravity, therefore no ghost
degrees of freedom occur in the theory.

\section{Conclusions}

The focus in this work was to enlighten the ghost problem of the
modified gravity theories containing the Gauss-Bonnet scalar
$\mathcal{G}$. Particularly, we studied two kind of theories,
namely $f(\mathcal{G})$ gravity and $F(R,\mathcal{G})$ gravity. In
both cases we investigated how the ghost degrees of freedom may
appear even at the equations of motion level, by using
perturbations of the metric, and as we demonstrated, ghost degrees
of freedom haunt both the aforementioned modified gravity
theories. In both cases, we provided a theoretical remedy by using
the Lagrange multiplier formalism which materializes constraints
in terms of the Lagrange multipliers. As we demonstrated, our
formalism leads to the elimination of the ghost degrees of freedom
in both the $f(\mathcal{G})$ gravity and $F(R,\mathcal{G})$
gravity theories, and thus the resulting theories can in principle
produce ghost free primordial curvature perturbations. Especially,
in the $F(R,\mathcal{G})$ gravity case, this was a serious issue
due to the fact that modes $\sim k^4$ occurred in the master
equation which governed the evolution of the primordial curvature
perturbations. For the case of the ghost-free $f(\mathcal{G})$
gravity theory, we investigated how accelerating cosmologies can
be realized by these theories. The formalism which we presented
can be used as a reconstruction technique, and as we demonstrated
there is room for rich model building, since in principle any
cosmological evolution can be realized by a number of different
ghost-free $f(\mathcal{G})$ theories, due to the freedom provided
by the Lagrange multiplier formalism. A future step of the results
we presented, is to provide a concrete formalism to study the
 inflationary period which can be technically difficult, due to the presence of the Lagrange multiplier. Work is
in progress towards this research line.

\section*{Acknowledgments}

This work is supported by MINECO (Spain), FIS2016-76363-P, and by
project 2017 SGR247 (AGAUR, Catalonia) (S.D.O). This work is also
supported by MEXT KAKENHI Grant-in-Aid for Scientific Research on
Innovative Areas ``Cosmic Acceleration'' No. 15H05890 (S.N.) and
the JSPS Grant-in-Aid for Scientific Research (C) No. 18K03615
(S.N.).

\end{document}